\documentclass[journal,twoside,web]{ieeecolor}
\usepackage{generic}
\usepackage{cite}
\usepackage{amsmath,amssymb,amsfonts}
\usepackage{graphicx}
\usepackage{algorithm,algorithmic}
\usepackage{hyperref}
\hypersetup{hidelinks=true}
\usepackage{textcomp}
\usepackage{cleveref}
\usepackage{booktabs}
\usepackage{colortbl}
\usepackage[caption=false,font=footnotesize]{subfig}   
\usepackage{enumerate}   
\usepackage{multirow}
\usepackage{tikz}
\usepackage{lipsum}
\usepackage[most]{tcolorbox}
\newtcolorbox{examplebox}[1][]{breakable,enhanced,colback=blue!4!white,colframe=blue!35!white,boxrule=0.4pt,arc=1.5pt,left=4pt,right=4pt,top=3pt,bottom=3pt,before skip=6pt,after skip=6pt,fontupper=\small,#1}
\newcounter{example}
\crefname{example}{Example}{Examples}
\newcommand{\exlabel}[2]{\refstepcounter{example}\label{#2}\textbf{Example~\theexample\ (#1).}~}
\usetikzlibrary{arrows.meta,calc}

\newtheorem{theorem}{Theorem}
\newtheorem{definition}{Definition}
\newtheorem{proposition}{Proposition}
\newtheorem{lemma}{Lemma}
\newtheorem{corollary}{Corollary}
\newtheorem{assumption}{Assumption}
\newtheorem{remark}{Remark}

\crefname{equation}{}{}
\crefname{section}{Sec.}{Sections}
\crefname{table}{Tab.}{Tables}
\crefname{algorithm}{Alg.}{Algorithms}
\crefname{definition}{Def.}{Def.}
\crefname{figure}{Fig.}{Figures}
\crefname{problem}{Prob.}{Problems}
\crefname{proposition}{Prop.}{Propositions}
\crefname{theorem}{Thm.}{Theorems}
\crefname{lemma}{Lem.}{Lemmas}
\crefname{remark}{Remark}{Remarks}
\crefname{corollary}{Cor.}{Corollaries}
\crefname{modelassumption}{Model Class}{Model Classes}
\crefname{assumption}{Assumption}{Assumptions}
\crefname{line}{Line}{Lines}

\newcommand{\X}{\mathcal{X}}
\newcommand{\Z}{\mathcal{Z}}
\newcommand{\XO}{\mathcal{X}_{\Delta,0}}

\newcommand{\Gx}{G_{\mathrm{x}}}
\newcommand{\Gu}{G_{\mathrm{u}}}
\newcommand{\ax}{\alpha_{\mathrm{x}}}
\newcommand{\au}{\alpha_{\mathrm{u}}}
\newcommand{\nv}{n_{\mathrm{v}}}
\newcommand{\nux}{\nu_{\mathrm{x}}}
\newcommand{\nuu}{\nu_{\mathrm{u}}}

\newcommand{\bx}{\beta_{\mathrm{x}}}
\newcommand{\bu}[1]{\beta_{\mathrm{u},#1}}
\newcommand{\lambx}{\lambda_{\mathrm{x}}}
\newcommand{\lambu}[1]{\lambda_{\mathrm{u},#1}}
\newcommand{\nm}{n_{\mathrm{m}}}
\newcommand{\ncal}{n_{\mathrm{cal}}}

\newcommand{\nx}{n_{\mathrm{x}}}
\newcommand{\nin}{n_{\mathrm{u}}}
\newcommand{\ny}{n_{\mathrm{y}}}

\newcommand{\nk}{n_{\mathrm{k}}}

\newcommand{\ntheta}{d}

\newcommand{\sys}{S}
\newcommand{\sysT}{S_{\mathrm{T}}}
\newcommand{\sysM}{S_{\mathrm{M}}}

\newcommand{\Y}{\mathcal{Y}}
\newcommand{\M}{\mathcal{M}}

\newcommand{\Meval}{\mathcal{M}_{\mathrm{eval}}}

\DeclareMathOperator*{\argmin}{arg\,min}

\newcommand{\Px}{\Pi_{\mathrm{x}}}
\newcommand{\Py}{\Pi_{\mathrm{y}}}
\newcommand{\PT}{\Pi_{\mathcal{T}}}

\newcommand{\Aprop}[2]{\bar{A}_{#1}^{(#2)}}
\newcommand{\Bprop}[3]{\bar{B}_{#1,#2}^{(#3)}}
\newcommand{\Cprop}[3]{\bar{C}_{#1,#2}^{(#3)}}
\newcommand{\Dprop}{\bar{D}}

\newcommand{\Apropm}[1]{\bar{A}_{#1}}
\newcommand{\Bpropm}[2]{\bar{B}_{#1,#2}}
\newcommand{\Cpropm}[2]{\bar{C}_{#1,#2}}
\newcommand{\ATpropm}{\bar{A}_{\mathcal{T}}}
\newcommand{\BTpropm}{\bar{B}_{\mathcal{T}}}

\allowdisplaybreaks[4]

\def\BibTeX{{\rm B\kern-.05em{\sc i\kern-.025em b}\kern-.08em
    T\kern-.1667em\lower.7ex\hbox{E}\kern-.125emX}}
\begin{document}
\title{Data-Driven Identification of Uncertainty Sets with Conformance Guarantees}
\author{
Laura L\"utzow, Mykel J. Kochenderfer, and Matthias Althoff
\thanks{This work has been submitted to the IEEE for possible publication. Copyright may be transferred without notice, after which this version may no longer be accessible.}
\thanks{L. L\"utzow and M. Kochenderfer are both with the Department of Aeronautics and Astronautics, Stanford University, USA (e-mail: {\tt\small lauralue@stanford.edu}).}
\thanks{M. Althoff is with the School of Computation, Information and Technology,
Technical University of Munich, Germany.}}

\maketitle

\begin{abstract}
Formal verification and controller synthesis require accurate models. Conformance relations specify the required accuracy of a model such that properties established for the model carry over to the target system. Existing methods for establishing them from measured data are typically restricted to a single relation, a specific model class, or a downstream verification algorithm. We present ConfID, a unified framework that establishes reachset, trace, and simulation conformance with respect to measured traces by identifying uncertainty sets for a given, possibly nonlinear nominal model. Unmodeled effects are thus captured by nondeterminism rather than by refining the nominal dynamics. For all three conformance relations, the identification is a linear program that minimizes the size of the reachable sets; only its constraints depend on the relation. We further derive distribution-free bounds on the probability that the identified models are conformant for unseen behaviors.
Experiments on simulated and real-world data evaluate the models identified for the three relations, confirm the guarantees, and show that these models cover unseen behaviors more reliably than baselines, with comparable or smaller sets.
\end{abstract}

\tikzset{%
  rset/.style={fill=black!12,draw=black!35,line width=0.25pt},
  mtrace/.style={draw=blue!65!black,line width=0.3pt},
  measline/.style={draw=red!65,line width=0.9pt,dotted},
  measdot/.style={circle,draw=red!80!black,line width=0.4pt,inner sep=0pt,minimum size=2.1pt},
  ax/.style={-{Stealth[length=3pt]},line width=0.3pt,black!70},
  tick/.style={font=\footnotesize,black!70,inner sep=0.6pt},
  albl/.style={font=\footnotesize},
  leglbl/.style={font=\footnotesize},
  nlbl/.style={font=\footnotesize,red!60,align=center}}
\newcommand{\confaxis}[1]{%
  \draw[ax] (-0.28,0.1) -- (-0.28,2.25);
  \draw[ax] (-0.28,0.1) -- (4.4,0.1);
  \foreach \k in {0,1,2,3,4} {\draw[black!70,line width=0.3pt] (\k,0.1) -- (\k,0.0) node[below,tick] {\k};}
  \node[tick,anchor=west] at (4.4,0.1) {$k$};
  \node[albl,rotate=90,anchor=south] at (-0.62,1.1) {#1};}
\newcommand{\rbars}[1]{\foreach \k/\lo/\hi in {#1} {\fill[rset] (\k-0.16,\lo) rectangle (\k+0.16,\hi);}}

\section{Introduction}

Formal verification and correct-by-construction controller synthesis provide rigorous guarantees that a dynamical system satisfies a given specification, but they require a model that captures every behavior that the system can exhibit. Deterministic models obtained from first principles or from classical system identification do not provide this property due to the variability of the real world. They are fit so as to minimize an average prediction error and therefore describe the system well on average, but they do not bound the deviation of the real system from the model. Safety verification, however, requires such bounds, since a property verified on the model only carries over to the real system if all deviations between the two are accounted for.
 
\begin{figure}[t]
\centering
\definecolor{confReach}{RGB}{239,108,0}
\definecolor{confTrace}{RGB}{251,192,45}
\definecolor{confSim}{RGB}{27,94,32}
\begin{tikzpicture}[x=1cm,y=1cm]
\begin{scope}[shift={(0,0)}]
\fill[rset,fill=black!7,draw=black!25] (0,-0.09) rectangle (0.16,0.09);
\node[leglbl,anchor=west,inner sep=0pt] (confLegA0) at (0.240,0) {Reachable sets of $\sysM$};
\end{scope}
\begin{scope}[shift={([xshift=0.4cm]confLegA0.east)}]
\draw[measline,draw=black!85,line width=0.9pt] (0,0) -- (0.4,0);
\node[measdot,draw=black!85,fill=white,minimum size=2.6pt] at (0.2,0) {};
\node[leglbl,anchor=west,inner sep=0pt] (confLegA1) at (0.480,0) {Measurements of $\sysT$};
\end{scope}
\end{tikzpicture}
\\[-1pt]
\begin{tikzpicture}[x=1cm,y=1cm]
\begin{scope}[shift={(0,0)}]
\draw[mtrace,draw=black!35,line width=0.6pt] (0,0) -- (0.4,0);
\node[leglbl,anchor=west,inner sep=0pt] (confLegB0) at (0.480,0) {Traces of $\sysM$};
\end{scope}
\begin{scope}[shift={([xshift=0.4cm]confLegB0.east)}]
\draw[mtrace,draw=black!55,line width=1.6pt] (0,0) -- (0.4,0);
\node[leglbl,anchor=west,inner sep=0pt] (confLegB1) at (0.480,0) {Trace of $\sysM$ reproducing the measurements};
\end{scope}
\end{tikzpicture}

\vspace{-0.3cm}

\subfloat[Reachset.]{\label{fig:confReach}%
\begin{tikzpicture}[x=0.510cm,y=0.60cm,rset/.append style={fill=confReach!12,draw=confReach!40},mtrace/.append style={draw=confReach!70,line width=0.6pt},wtrace/.style={mtrace,draw=confReach,line width=1.6pt},measline/.append style={draw=black!85,line width=0.9pt},measdot/.append style={draw=black!85,fill=white,minimum size=2.6pt},nlbl/.append style={black!55}]
  \confaxis{}
  \rbars{0/1.362/1.415,1/0.468/0.587,2/0.580/0.779,3/1.079/1.371,4/0.560/0.953}
  \draw[mtrace] (0,1.389) -- (1,0.527) -- (2,0.680) -- (3,1.225) -- (4,0.756);
  \draw[mtrace] (0,1.362) -- (1,0.468) -- (2,0.580) -- (3,1.079) -- (4,0.560);
  \draw[mtrace] (0,1.415) -- (1,0.587) -- (2,0.779) -- (3,1.371) -- (4,0.953);
  \node[nlbl] at (2.050,1.850) {not measured};
  \draw[black!70,line width=0.3pt] (-0.360,0.561) -- (-0.280,0.561);
  \draw[black!70,line width=0.3pt] (-0.360,1.305) -- (-0.280,1.305);
  \begin{scope}[yshift=-1.70cm]
    \confaxis{}
    \rbars{0/0.518/0.853,1/0.801/1.229,2/1.005/1.518,3/1.080/1.667,4/1.184/1.833}
    \draw[mtrace] (0,0.686) -- (1,1.015) -- (2,1.262) -- (3,1.373) -- (4,1.508);
    \draw[mtrace] (0,0.853) -- (1,1.229) -- (2,1.518) -- (3,1.667) -- (4,1.833);
    \draw[mtrace] (0,0.518) -- (1,0.801) -- (2,1.005) -- (3,1.080) -- (4,1.184);
    \draw[measline] (0,0.808) -- (1,0.801) -- (2,1.438) -- (3,1.667) -- (4,1.264);
    \node[measdot] at (0,0.808) {};
    \node[measdot] at (1,0.801) {};
    \node[measdot] at (2,1.438) {};
    \node[measdot] at (3,1.667) {};
    \node[measdot] at (4,1.264) {};
    \draw[black!70,line width=0.3pt] (-0.360,1.350) -- (-0.280,1.350);
    \draw[black!70,line width=0.3pt] (-0.360,1.933) -- (-0.280,1.933);
  \end{scope}
  \coordinate (rowlbla) at (-0.360,1.175);
  \coordinate (rowlblb) at (-0.360,-1.658);
  \node[albl,inner sep=0pt,rotate=90,anchor=south] at ([xshift=-2pt]rowlbla) {State};
  \node[albl,inner sep=0pt,rotate=90,anchor=south] at ([xshift=-2pt]rowlblb) {Output};
\end{tikzpicture}}
\hfil
\subfloat[Trace.]{\label{fig:confTrace}%
\begin{tikzpicture}[x=0.478cm,y=0.60cm,rset/.append style={fill=confTrace!12,draw=confTrace!40},mtrace/.append style={draw=confTrace!85!black,line width=0.6pt},wtrace/.style={mtrace,draw=confTrace!80!black,line width=1.6pt},measline/.append style={draw=black!85,line width=0.9pt},measdot/.append style={draw=black!85,fill=white,minimum size=2.6pt},nlbl/.append style={black!55}]
  \confaxis{}
  \rbars{0/1.389/1.389,1/0.526/0.529,2/0.675/0.684,3/1.215/1.234,4/0.741/0.772}
  \draw[mtrace] (0,1.389) -- (1,0.527) -- (2,0.680) -- (3,1.225) -- (4,0.756);
  \draw[mtrace] (0,1.389) -- (1,0.526) -- (2,0.675) -- (3,1.215) -- (4,0.741);
  \draw[mtrace] (0,1.389) -- (1,0.529) -- (2,0.684) -- (3,1.234) -- (4,0.772);
  \draw[mtrace] (0,1.389) -- (1,0.526) -- (2,0.675) -- (3,1.215) -- (4,0.744);
  \node[nlbl] at (2.050,1.850) {not measured};
  \draw[black!70,line width=0.3pt] (-0.360,0.561) -- (-0.280,0.561);
  \draw[black!70,line width=0.3pt] (-0.360,1.305) -- (-0.280,1.305);
  \begin{scope}[yshift=-1.70cm]
    \confaxis{}
    \rbars{0/0.422/0.949,1/0.742/1.289,2/0.978/1.545,3/1.080/1.667,4/1.206/1.811}
    \draw[mtrace] (0,0.686) -- (1,1.015) -- (2,1.262) -- (3,1.373) -- (4,1.508);
    \draw[mtrace] (0,0.686) -- (1,1.025) -- (2,1.282) -- (3,1.403) -- (4,1.811);
    \draw[mtrace] (0,0.686) -- (1,1.005) -- (2,1.241) -- (3,1.343) -- (4,1.206);
    \draw[wtrace] (0,0.808) -- (1,0.801) -- (2,1.438) -- (3,1.667) -- (4,1.264);
    \draw[measline] (0,0.808) -- (1,0.801) -- (2,1.438) -- (3,1.667) -- (4,1.264);
    \node[measdot] at (0,0.808) {};
    \node[measdot] at (1,0.801) {};
    \node[measdot] at (2,1.438) {};
    \node[measdot] at (3,1.667) {};
    \node[measdot] at (4,1.264) {};
    \draw[black!70,line width=0.3pt] (-0.360,1.350) -- (-0.280,1.350);
    \draw[black!70,line width=0.3pt] (-0.360,1.933) -- (-0.280,1.933);
  \end{scope}
\end{tikzpicture}}
\hfil
\subfloat[Simulation.]{\label{fig:confSim}%
\begin{tikzpicture}[x=0.478cm,y=0.60cm,rset/.append style={fill=confSim!12,draw=confSim!40},mtrace/.append style={draw=confSim!55,line width=0.6pt},wtrace/.style={mtrace,draw=confSim!75,line width=1.6pt},measline/.append style={draw=black!85,line width=0.9pt},measdot/.append style={draw=black!85,fill=white,minimum size=2.6pt},nlbl/.append style={black!55}]
  \confaxis{}
  \rbars{0/1.306/1.471,1/0.361/0.694,2/0.429/0.930,3/0.893/1.557,4/0.347/1.166}
  \draw[mtrace] (0,1.389) -- (1,0.527) -- (2,0.680) -- (3,1.225) -- (4,0.756);
  \draw[mtrace] (0,1.471) -- (1,0.690) -- (2,0.919) -- (3,1.453) -- (4,0.965);
  \draw[mtrace] (0,1.306) -- (1,0.364) -- (2,0.440) -- (3,0.997) -- (4,0.547);
  \draw[wtrace] (0,1.365) -- (1,0.464) -- (2,0.531) -- (3,1.086) -- (4,0.674);
  \draw[measline] (0,1.365) -- (1,0.464) -- (2,0.531) -- (3,1.086) -- (4,0.674);
  \node[measdot] at (0,1.365) {};
  \node[measdot] at (1,0.464) {};
  \node[measdot] at (2,0.531) {};
  \node[measdot] at (3,1.086) {};
  \node[measdot] at (4,0.674) {};
  \draw[black!70,line width=0.3pt] (-0.360,0.561) -- (-0.280,0.561);
  \draw[black!70,line width=0.3pt] (-0.360,1.305) -- (-0.280,1.305);
  \begin{scope}[yshift=-1.70cm]
    \confaxis{}
    \rbars{0/0.403/0.968,1/0.721/1.309,2/0.943/1.580,3/1.018/1.728,4/1.104/1.913}
    \draw[mtrace] (0,0.686) -- (1,1.015) -- (2,1.262) -- (3,1.373) -- (4,1.508);
    \draw[mtrace] (0,0.686) -- (1,1.027) -- (2,1.298) -- (3,1.446) -- (4,1.913);
    \draw[mtrace] (0,0.686) -- (1,1.004) -- (2,1.226) -- (3,1.300) -- (4,1.104);
    \draw[wtrace] (0,0.808) -- (1,0.801) -- (2,1.438) -- (3,1.667) -- (4,1.264);
    \draw[measline] (0,0.808) -- (1,0.801) -- (2,1.438) -- (3,1.667) -- (4,1.264);
    \node[measdot] at (0,0.808) {};
    \node[measdot] at (1,0.801) {};
    \node[measdot] at (2,1.438) {};
    \node[measdot] at (3,1.667) {};
    \node[measdot] at (4,1.264) {};
    \draw[black!70,line width=0.3pt] (-0.360,1.350) -- (-0.280,1.350);
    \draw[black!70,line width=0.3pt] (-0.360,1.933) -- (-0.280,1.933);
  \end{scope}
\end{tikzpicture}}
\caption{Individual models $\sysM$ identified for the three conformance relations from one measured trajectory of a pendulum (\cref{ex:pendulum}). For simplicity, we plot only the second state component $x_{k+1(2)}$ (angular velocity) at the top and the output $y_k$ (measured angle) at the bottom. (a) Reachset conformance requires each measured output to lie in the reachable set; the measured sequence need not be a trace of the model. (b) Trace conformance requires that the model reproduces each output trace. (c) Simulation conformance requires state measurements and that the model reproduces the traces of the state and the output.
}
\label{fig:conformanceOverview}
\vspace{-0.3cm}
\end{figure}
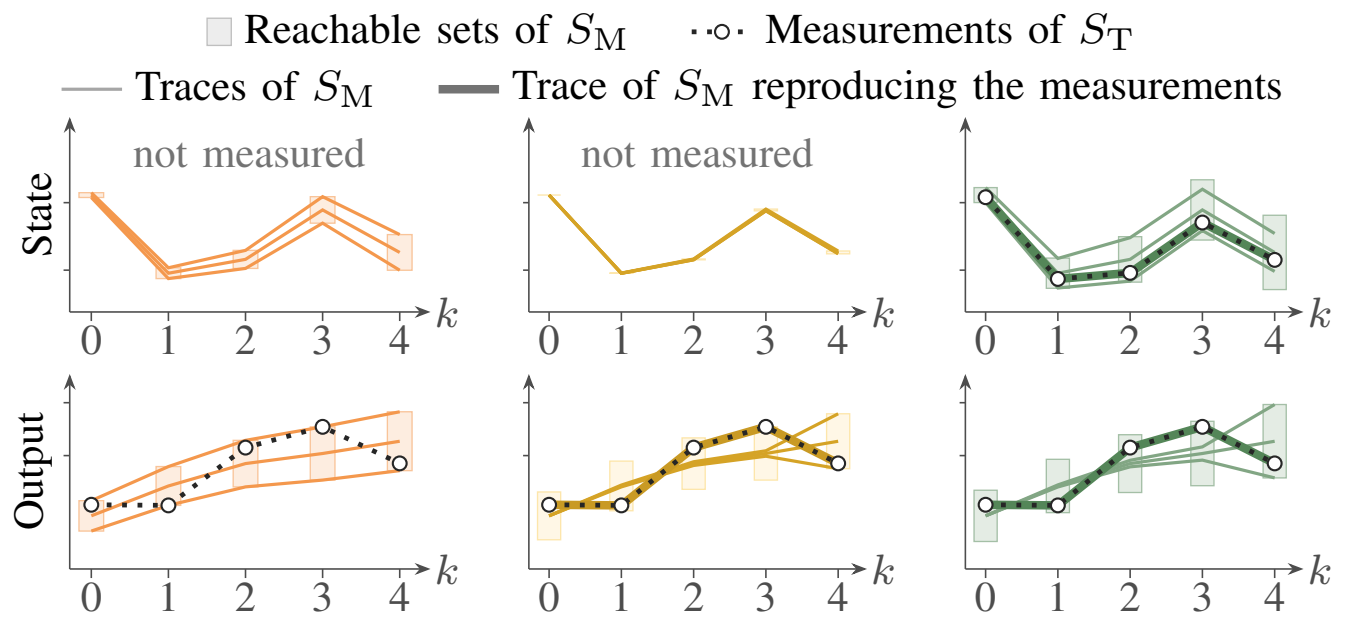

Abstraction synthesis addresses this issue by constructing reduced-complexity models, referred to as abstractions, that preserve selected properties of the target system. Such reductions are typically achieved through state and input discretization or by introducing controlled nondeterminism, which additionally renders formal verification and controller synthesis tractable \cite{kazemi2014abstractionControl}. The class of properties that can be transferred from an abstraction to the target system is determined by the underlying behavioral relation \cite{roehm2019conformance,tabuada2009symbolic}.
Stricter relations transfer richer specifications, from safety properties verified by reachability analysis \cite{roehm2016reachset} over temporal logic specifications on outputs \cite{roehm2019conformance} to state-space properties \cite{tabuada2009symbolic,sadraddini2018formal}, but they impose more restrictive requirements on the abstraction and generally increase its conservatism. 
Existing identification methods, reviewed in \cref{sec:relWork}, target a single relation and are mostly confined to linear dynamics, finite abstractions, or a specific reachability algorithm, so that changing the relation requires changing the method, the model class, and the tooling.

This paper presents ConfID, a framework for establishing reachset conformance, trace conformance, and simulation conformance (\cref{fig:conformanceOverview}) over continuous state and input spaces, from measured trajectories of the target system. We take a nominal model as given, e.g., obtained from first principles or classical system identification, and identify the uncertainty sets that establish conformance, where the uncertainty sets bound the deviations of the initial state and the inputs as well as additive uncertainties in the state and output equations.
These sets must be large enough to cover all behaviors of the target system, yet as small as possible to limit the behaviors that the model admits but the target system cannot exhibit. 
Since unmodeled effects are captured by nondeterminism rather than by refining the nominal dynamics, the physical interpretability of the nominal model can be preserved. Under a zonotopic parametrization, the identification is reduced to a linear program that minimizes a surrogate measure of conservatism and whose result carries distribution-free guarantees for unseen behaviors. The three identification problems differ only in the containment constraints on the measured trajectories, whereas the model class, the parametrization, and the solver remain unchanged.

This paper presents the following contributions:
\begin{enumerate}
    \item We identify uncertainty sets that render a possibly nonlinear model reachset-conformant, trace-conformant, or simulation-conformant with respect to the target system for a given set of measured trajectories. We show that all three problems reduce to linear programs, in which the stricter relation trace conformance requires fewer auxiliary variables than reachset conformance.
    \item We bound the effective dimension of these programs and derive conformance guarantees for unseen trajectories.
    \item We compare the three relations on two simulated systems and on measured data of an industrial robot, and against split conformal calibration and a residual-box baseline.
\end{enumerate}
The remainder of this paper is organized as follows. \cref{sec:relWork} reviews related work, \cref{sec:prelim} introduces the three conformance relations and states the identification problem, \cref{sec:framework} develops our identification framework, \cref{sec:guarantees} derives generalization guarantees for unseen trajectories, and \cref{sec:results} evaluates the approach numerically.

\section{Related Work} \label{sec:relWork}

This section reviews existing data-driven methods for establishing reachset conformance, trace conformance, and simulation conformance, ordered from the weakest to the strongest relation. For each relation, we summarize the properties it preserves, review existing identification approaches, and contrast them with our framework.

\subsection{Reachset-Conformant Identification}

Reachset conformance requires that, at each time step, all outputs of the target system are contained within the reachable set of the abstraction \cite{roehm2016reachset}. Consequently, any output deemed unreachable by the abstraction is guaranteed to be unreachable for the true system. This property enables the sound transfer of safety specifications verified through reachability analysis, as well as temporal properties expressed in reachset temporal logic \cite{roehm2016stl}.
Recent compositionality results further show that reachset-conformant subsystem models can be combined into a reachset-conformant model of the overall system, thereby supporting scalable analysis of complex interconnected systems \cite{luetzow2026prod}.
Due to its favorable trade-off between computational tractability and formal guarantees, reachset conformance has gained significant attention in safety verification \cite{althoff2019effortless,schaefer2024scalable,walter2025leveraging,althoff2025traffic}.

Reachset-conformant identification is related to data-driven reachability analysis \cite{devonport2021data,alanwar2023data,tebjou2023data,hashemi2023data,park2024data,dietrich2025data}; however, the latter approximates reachable sets directly from data, whereas the former identifies a model whose reachable sets are consistent with the data, which enables model-based verification, controller synthesis, and other downstream tasks. Existing methods bound prediction errors by intervals using state measurements \cite{kochdumper2020conf}, compute uncertainty sets for linear \cite{gruber2023scalable,Liu2023conf,althoff2023conf,luetzow2024generator}, hybrid \cite{tang2024efficiently}, and nonlinear models \cite{althoff2012reachability,luetzow2026reachset}, incorporate new data recursively \cite{luetzow2025recursive}, and bound the violation probability for unseen data by scenario theory \cite{luetzow2026prod,calafiore2006scenario}.

In contrast to existing approaches, our method is not restricted to linear model classes or specific reachability algorithms, and can be applied to any twice continuously differentiable model while retaining formal guarantees.

\subsection{Trace-Conformant Identification}

Trace conformance requires that every output trajectory of the target system can be reproduced by the abstraction \cite{roehm2019conformance,tabuada2009symbolic}. This notion appears under various terminologies in the literature, including behavioral inclusion\cite{tabuada2009symbolic,coppola2023dataAbstractionLinear}, language inclusion~\cite{alur2001compositionalRefine,ouaknine2004language}, refinements~\cite{henzinger2001assume}, implementation relation~\cite{lynch2001hybrid}, or hybrid input-output conformance~\cite{vanOsch2006hioco}. 
As it enforces trajectory-wise consistency, trace conformance is strictly stronger than reachset conformance and typically necessitates richer model structures or increased nondeterminism. It ensures preservation of safety properties defined over output trajectories.

Most existing works construct trace-conformant abstractions with finite state and output spaces \cite{coppola2023dataAbstractionLinear,peruffo2023data,peruffo2025sampling,banse2025dataMemoryAbstraction,banse2023dataRefinement,banse2023dataMemeory}, often with probably approximately correct guarantees from scenario theory \cite{peruffo2023data,peruffo2025sampling}. For continuous-state systems, deterministic models are augmented with uncertainty, and conformance is established by refining the model or enlarging the uncertainty until a constraint satisfaction problem over the uncertainty realizations becomes feasible \cite{schuermann2017ensuring} or by computing the minimal additive disturbances that reproduce the measured trajectories with a linear program \cite{kochdumper2022conformant}. 

Compared to existing methods, our approach allows for significantly greater flexibility in both model structure and uncertainty representation, which can reduce conservatism while still providing probabilistic guarantees for unseen data.

\subsection{Simulation-Conformant Identification}

Simulation relations, the strongest relations considered in this work, require that every trajectory of the target system admits a state-wise corresponding trajectory of the abstraction with identical outputs. A bidirectional simulation relation yields bisimilarity, which guarantees preservation of expressive temporal logic specifications \cite{priser2024robust}. With inputs, simulation relations can be formulated under identical input sequences~\cite{roehm2019conformance} or as feedback refinement relations, which enable correct-by-construction controller synthesis \cite{reissig2014feedback}. Alternating simulation relaxes input matching requirements, while $\epsilon$-approximate simulation allows bounded output deviations \cite{tabuada2009symbolic}. For stochastic systems, approximate probabilistic bisimulations bound the mismatch between the transition probabilities of two Markov processes~\cite{abate2013approximation} and imply approximate trace equivalence~\cite{bian2017relationship}.

Data-driven identification of simulation-conformant models typically requires full state measurements and mostly relies on state-space discretization: alternating (bi)simulation relations have been established for monotone systems and their disturbed variants \cite{makdesi2021data,makdesi2021efficient}, finite abstractions satisfying $\epsilon$-approximate alternating simulation have been constructed with probably approximately correct statistics \cite{devonport2021symbolic} or Gaussian processes \cite{hashimoto2022learning}, and further approaches use stochastic or alternating bisimulation functions from robust optimization \cite{lavaei2023data,lavaei2023constructing}, counterexample-guided refinement \cite{abate2024bisimulation}, or non-uniform gridding and data-driven partitioning \cite{dinnocenzo2025data,ajeleye2023data}, often with scenario-based guarantees \cite{lavaei2023data,lavaei2023constructing,dinnocenzo2025data}. To the best of the authors' knowledge, the only work over continuous state spaces fits piecewise affine dynamics by mixed-integer programming and then optimizes hyper-rectangular disturbance sets that bound the trajectory deviations \cite{sadraddini2018formal}.

In contrast, our approach accommodates nonlinear model classes and non-additive, structured uncertainty representations, thereby reducing conservatism while maintaining probabilistic guarantees for unseen data.

\section{Preliminaries} \label{sec:prelim}
\subsection{Notation}
Sets are denoted by calligraphic letters, matrices by uppercase letters, and vectors and scalars by lowercase letters. The symbols $\mathbf{1}$, $\mathbf{I}$, and $\mathbf{0}$ denote, respectively, an all-ones vector, an identity matrix, and a zero matrix of appropriate dimensions; dimensions are specified through subscripts when needed.

For a vector $v\in\mathbb{R}^n$, $\mathrm{diag}(v)\in\mathbb{R}^{n\times n}$ denotes the diagonal matrix with entries of $v$ on its main diagonal. For the vectors $v_1,\dots,v_k\in\mathbb{R}^n$, we define their vertical concatenation as $v_{1:k}=[v_1^\top~\dots ~v_k^\top]^\top\in\mathbb{R}^{kn}$. We denote the $i$-th component of a vector $v$ by $v_{(i)}$ and its components $i$ to $j$ by $v_{(i:j)}$. 
For the matrices $A_1,\dots,A_k\in\mathbb{R}^{m\times n}$, we define their vertical concatenation as $A_{1:k}=[A_1^\top~\dots ~A_k^\top]^\top\in\mathbb{R}^{km\times n}$.
The operator $\mathrm{vert}_{k}(A)\in\mathbb{R}^{km\times n}$ denotes the vertical concatenation of $k$ copies of the matrix $A\in\mathbb{R}^{m\times n}$, and $\mathrm{diag}(A_1,\dots,A_k)$ denotes the block-diagonal matrix formed by the matrices $A_1,\dots,A_k$. 

For a set $\mathcal{S}\subset\mathbb{R}^n$ and a matrix $A\in\mathbb{R}^{m\times n}$, the image of $\mathcal{S}$ under the linear map $A$ is defined as $A\mathcal{S} := \{As \mid s\in\mathcal{S}\}\subseteq\mathbb{R}^m$. We write $-\mathcal{S} := (-1)\mathcal{S}$ and, for a vector $v\in\mathbb{R}^n$, $\mathcal{S}-v := \mathcal{S}\oplus\{-v\}$. For sets $\mathcal{S}_1,\mathcal{S}_2\subset\mathbb{R}^n$, the Minkowski sum is defined as $\mathcal{S}_1 \oplus \mathcal{S}_2 := \{s_1 + s_2 \mid s_1\in\mathcal{S}_1,\ s_2\in\mathcal{S}_2\}\subset\mathbb{R}^n$. 
For sets $\mathcal{S}_1\subset\mathbb{R}^{n_1},~\mathcal{S}_2\subset\mathbb{R}^{n_2}$, the Cartesian product is defined as $\mathcal{S}_1 \times \mathcal{S}_2 := \{[s_1^\top\ s_2^\top]^\top \mid s_1\in\mathcal{S}_1,\ s_2\in\mathcal{S}_2\}\subset\mathbb{R}^{n_1+n_2}$.
We denote the Cartesian product of the sets $\mathcal{S}_1,\dots,\mathcal{S}_k\subset\mathbb{R}^{n}$ as $\mathcal{S}_{1:k}=\mathcal{S}_1\times\dots\times\mathcal{S}_k\subset\mathbb{R}^{kn}$. The decorations used throughout the paper are summarized in \cref{tab:notation}.
\begin{table}[t]
\centering
\caption{Recurring decorations and symbols.}
\label{tab:notation}
\footnotesize
\begin{tabular}{@{}lp{5.2cm}@{}}
\toprule
Symbol & Meaning; where defined\\
\midrule
$\bar{x}_k,\bar{u}_k,\bar{y}_k,\bar{z}_k$ & nominal or reference quantity; \cref{def:testcase}, \cref{eq:refTraj}\\
$x_{\Delta,k},u_{\Delta,k},y_{\Delta,k},z_{\Delta,k}$ & deviation from the nominal quantity, e.g., $u_{\Delta,k}=u_k-\bar u_k$; \cref{eq:taylor}, \cref{prop:constrAll}\\
$\XO,\mathcal{U}_{\Delta,k}$ & admissible deviation sets, scaled by $\alpha$; \cref{eq:uncSetsStructure}\\
$e_{\mathrm{x},k},e_{\mathrm{y},k}\in\mathcal{E}_{\mathrm{x},k},\mathcal{E}_{\mathrm{y},k}$ & additive model uncertainty; \cref{eq:sysNL}, \cref{eq:uncChoice}\\
${\X}_k,{\Y}_k,{\Z}_k$ & reachable set; \cref{def:reach}\\
${\mathcal{T}}_{\mathrm{x}},{\mathcal{T}}_{\mathrm{y}},{\mathcal{T}}_{\mathrm{z}}$ & trace set; \cref{def:trace}\\
$\widehat{\X}_k,\widehat{\Y}_k,\widehat{\Z}_k,\widehat{\mathcal{L}}$ & tractable overapproximation; \cref{thm:jointReach}, \cref{eq:LHat}\\
$\bar{\Y}_k,\bar{\Z}_k,\bar{\mathcal{T}}_{\mathrm{y}},\bar{\mathcal{T}}_{\mathrm{z}}$ & linearization-induced set; \cref{eq:ZBar}, \cref{eq:TzBar}, \cref{eq:linObjects}\\
$\Apropm{k},\Bpropm{k}{i},\Cpropm{k}{i},\Dprop$ & propagation matrices of the linearization; \cref{eq:propMat}\\
$\ATpropm,\BTpropm$ & stacked propagation matrices; \cref{eq:propMatTrace}\\
$(\cdot)^{(m)}$, $(\cdot)^{(m,k)}$ & test case $m$ and one-step test case starting at $x_k^{(m)}$; \cref{def:testcase}, \cref{sec:confData}\\
$\alpha,\beta,\gamma$ & scaling factors, auxiliary variables, cost vector; \cref{eq:uncSetsStructure}, \cref{prop:constrAll}, \cref{eq:costCoeff}\\
\bottomrule
\end{tabular}
\end{table}

\subsection{Conformance Relations}

Conformance relations formalize the correspondence between a target system $\sysT$ and an abstract model $\sysM$. Since the behavior of $\sysT$ is often only partially known or computationally intractable to analyze directly, conformance relations enable properties verified on the more tractable model $\sysM$ to be transferred to $\sysT$.

Consider a system $S$ with state $x_k \in \mathbb{R}^{\nx}$, input $u_{k} \in \mathbb{R}^{\nin}$, which comprises control inputs and disturbances, and output $y_k \in \mathbb{R}^{\ny}$ at time step $k$. The successor state $x_{k+1}$ and the output $y_k$ depend on the input $u_k$ and the current state $x_k$ and are collected in the vector $z_k = \bigl[x_{k+1}^\top~~y_k^\top\bigr]^\top \in \mathbb{R}^{\nx+\ny}$. 

In practice, the true initial state $x_0$ and input sequence ${u}_{0:k}$ of real-world systems are often not known exactly. Instead, nominal quantities $\bar{x}_0$ and $\bar{u}_{0:k}$ are available, e.g., obtained from state estimation and known control inputs, with disturbance components set to their expected values. Due to uncertainty and possible nondeterminism, identical nominal quantities may give rise to different state and output trajectories, hereafter called traces.

\begin{definition}[Trace set]\label{def:trace}
    The state trace set $\mathcal{T}_{\mathrm{x}}(S,\bar{x}_0,\bar{u}_{0:k})$, the output trace set $\mathcal{T}_{\mathrm{y}}(S,\bar{x}_0,\bar{u}_{0:k})$, and the joint trace set $\mathcal{T}_{\mathrm{z}}(S,\bar{x}_0,\bar{u}_{0:k})$ are the sets of all traces $x_{1:k+1}$, $y_{0:k}$, and $z_{0:k}$, respectively, that $\sys$ can produce for the nominal initial state $\bar{x}_0$ and input sequence $\bar{u}_{0:k}$ under all admissible uncertainties.
\end{definition}\vspace{0.2cm}

\begin{definition}[Reachable set]\label{def:reach}
The reachable state set $\X_k(S,\bar{x}_0,\bar{u}_{0:k})$, the reachable output set $\Y_k(S,\bar{x}_0,\bar{u}_{0:k})$, and the joint reachable set $\Z_k(S,\bar{x}_0,\bar{u}_{0:k})$ are the sets of all states $x_k$, outputs $y_k$, and stacked vectors $z_k$, respectively, that $\sys$ can attain for the nominal initial state $\bar{x}_0$ and input sequence $\bar{u}_{0:k}$ under all admissible uncertainties.
\end{definition}\vspace{0.2cm}

Reachable sets are defined per time step and do not preserve temporal correlations, so the Cartesian product $\mathcal{Y}_{0:k}(S,\bar{x}_0,\bar{u}_{0:k})$ of the reachable sets contains every trace in $\mathcal{T}_{\mathrm{y}}(S,\bar{x}_0,\bar{u}_{0:k})$ but generally also spurious traces whose transitions between consecutive steps cannot be realized by $\sys$ (see \cref{ex:pendConf}). 
Similarly, the joint reachable set $\Z_k$ is generally a strict subset of $\X_{k+1}\times\Y_k$, because the successor state and the output depend on the same input, whereas the Cartesian product does not preserve such dependencies.
For notational compactness, we group nominal data into test cases.

\begin{definition}[Test case]\label{def:testcase}
A test case $m$ of length $\nk$ is defined by a nominal initial state $\bar{x}_0^{(m)}$ and a nominal input sequence $\bar{u}_{0:\nk-1}^{(m)}$. All test cases under which the target system can be operated are collected in the test case space $\mathbb{M}$.
\end{definition}\vspace{0.2cm}

Since $\sysT$ is nondeterministic, the trace it produces in a test case is random as well; we let $\mathbb{P}$ denote the joint distribution of a test case and the trace measured when executing it.
The trace sets and 
reachable sets of system $\sys$ for test case $m$ are written as $\mathcal{T}_{\mathrm{x}}^{(m)}(\sys)$, 
$\mathcal{T}_{\mathrm{y}}^{(m)}(\sys)$, $\mathcal{T}_{\mathrm{z}}^{(m)}(\sys)$, $\X_k^{(m)}(\sys)$, 
$\Y_k^{(m)}(\sys)$, and $\Z_k^{(m)}(\sys)$.

\begin{examplebox}
\exlabel{Pendulum}{ex:pendulum} As a running example, consider a pendulum with angle $x_{(1)}$ and angular velocity $x_{(2)}$,
\begin{equation}\label{eq:pendulum}
x_{k+1}=x_k+\Delta t\begin{bmatrix}x_{k(2)}\\ -\tfrac{g}{l}\sin x_{k(1)}+\tau_k\end{bmatrix}+\begin{bmatrix}w_k\\ 0\end{bmatrix},
\end{equation}
with  step size $\Delta t$, output $y_k=x_{k(1)}+v_k$, and input $u=[\tau~w~v]^\top$ comprising the normalized torque $\tau$, the process disturbance $w$, and the measurement noise $v$. A test case consists of $\bar{x}_0=[\bar{x}_{0(1)}~\bar{x}_{0(2)}]^\top$ and $\bar{u}_k=[\bar{\tau}_k~0~0]^\top$, i.e., the planned normalized torques with the disturbance and the noise at their expected value zero. To see why $\Y_1\times\Y_2$ can contain spurious traces, let the pendulum rest at $\bar{x}_0=\mathbf{0}$ with $\bar{\tau}_k=0$, and let the only nonzero uncertainty be $w_0\in\{-a,a\}$. Since $x_{1(2)}=0$, we obtain $y_1=w_0$ and $y_2=y_1$, hence $\Y_1=\Y_2=\{-a,a\}$. Every admissible trace satisfies $y_2=y_1$, so $y_{1:2}=[a~{-a}]^\top\in\Y_1\times\Y_2$ is spurious: $y_2=-a$ requires $w_0=-a$, contradicting $y_1=a$.
\end{examplebox}

Based on these observations, we introduce three conformance relations of increasing strictness (\cref{fig:conformanceOverview}), assuming that $\sysT$ and $\sysM$ share the state, input, and output spaces.

\begin{definition}[{Reachset conformance \cite[Def. 6]{luetzow2026reachset}}] \label{def:reachConf}
    $\sysM$ is a reachset-conformant model of $\sysT$, written as $\sysT\preceq_\mathrm{reach}\sysM$, if the reachable set of $\sysM$ encloses the reachable set of $\sysT$ for all time steps $k$ and all test cases $m\in\mathbb{M}$:
    \begin{align*}
        \sysT\preceq_\mathrm{reach}\sysM \Leftrightarrow \forall k,\, m\in\mathbb{M}\colon \Y_k^{(m)}(\sysT) \subseteq \Y_k^{(m)}(\sysM).
    \end{align*}
\end{definition}\vspace{0.2cm}
Reachset conformance enables the transfer of output-space safety 
properties: if an output is not reachable under $\sysM$, it is not 
reachable under $\sysT$ either. More generally, any temporal property 
expressible in reachset temporal logic~\cite[Sec.~3.5]{roehm2019conformance} 
verified on $\sysM$ carries over to $\sysT$.

\begin{definition}[{Trace conformance \cite[Sec.~3.1]{roehm2019conformance}}] \label{def:traceConf}
    $\sysM$ is a trace-conformant model of $\sysT$, written as $\sysT\preceq_\mathrm{trace}\sysM$, if the set of all output traces of $\sysM$ contains all possible output traces of $\sysT$ for all test cases $m\in\mathbb{M}$:
    \begin{align*}
        \sysT\preceq_\mathrm{trace}\sysM \Leftrightarrow \forall m\in\mathbb{M}\colon \mathcal{T}_{\mathrm{y}}^{(m)}(\sysT) \subseteq \mathcal{T}_{\mathrm{y}}^{(m)}(\sysM).
    \end{align*}
\end{definition}\vspace{0.2cm}
Since the trace 
set $\mathcal{T}_{\mathrm{y}}^{(m)}$ preserves temporal correlations across time 
steps, trace conformance 
is strictly stronger than reachset conformance and additionally 
enables the transfer of linear temporal logic properties over the 
output space~\cite[Tab.~1]{roehm2019conformance}.

\begin{definition}[{Simulation conformance}] \label{def:simConf}
    System $\sysM$ is a simulation-conformant model of $\sysT$, written as $\sysT\preceq_\mathrm{sim}\sysM$, 
    if the set of joint traces of $\sysM$ contains all possible joint traces of $\sysT$ for all test cases $m\in\mathbb{M}$:
    \begin{align*}
        \sysT\preceq_\mathrm{sim}\sysM \Leftrightarrow \forall m\in\mathbb{M}\colon \mathcal{T}_{\mathrm{z}}^{(m)}(\sysT) \subseteq \mathcal{T}_{\mathrm{z}}^{(m)}(\sysM).
    \end{align*}
\end{definition}\vspace{0.2cm}
Simulation conformance means that $\sysM$ simulates $\sysT$ in the sense of a simulation relation. \cref{def:simConf} follows \cite[Def.~4.1]{sadraddini2018formal}, which, unlike other definitions, considers inputs and relates each state only to itself. It requires $\sysM$ to reproduce the states of $\sysT$ and thus relies on state measurements, which are available, e.g., when $\sysT$ is itself a high-fidelity model. Beyond~\cite[Def.~4.1]{sadraddini2018formal}, we also require $\sysM$ to reproduce the outputs of $\sysT$.
Since $\mathcal{T}_{\mathrm{z}}^{(m)}$ ties the state and output traces to a single uncertainty realization, the containment in \cref{def:simConf} is strictly stronger than requiring
$\mathcal{T}_{\mathrm{x}}^{(m)}(\sysT)\subseteq\mathcal{T}_{\mathrm{x}}^{(m)}(\sysM)$ and $\mathcal{T}_{\mathrm{y}}^{(m)}(\sysT)\subseteq\mathcal{T}_{\mathrm{y}}^{(m)}(\sysM)$. In addition to the output-space properties transferable with trace conformance, simulation conformance enables the transfer of state-space properties \cite[Tab.~1]{roehm2019conformance}, making it the most expressive---though also the most demanding---conformance criterion considered here.

\subsection{Zonotopes} \label{subsec:zon}

We represent sets by zonotopes, since they admit cheap computations and efficient representations in high dimensions. 

\begin{definition}[Zonotope~\cite{kuehn1998wrapping}] \label{def:zon}
A zonotope $\mathcal{S} \subset \mathbb{R}^n$ is defined as
\[
\mathcal{S} = \left\{ c + G\lambda \mid \lambda \in [-1,1]^{\nu} \right\} = \langle c, G \rangle,
\]
where $c \in \mathbb{R}^n$ is the center and $G \in \mathbb{R}^{n \times \nu}$ is the generator matrix.
\end{definition}\vspace{0.2cm}

Zonotopes are closed under linear maps, $A\langle c,G \rangle = \langle Ac, AG \rangle$, Minkowski sums, $\langle c_1,G_1 \rangle \oplus \langle c_2,G_2 \rangle = \langle c_1+c_2, [G_1~G_2] \rangle$, and Cartesian products, $\langle c_1,G_1 \rangle \times \langle c_2,G_2 \rangle=\langle [c_1^\top~c_2^\top]^\top,\mathrm{diag}(G_1, G_2)\rangle$.
A point $z \in \mathbb{R}^n$ belongs to $\langle c, G \rangle$ if and 
only if \cite[Eq.~(6)]{luetzow2024generator}
\begin{align}
    z \in \langle c,G \rangle~ \Leftrightarrow ~\exists\lambda \in \mathbb{R}^{\nu}\colon |\lambda| \leq \textbf{1},~ z = c+G\lambda. \label{eq:cont_gen}
\end{align}

To quantify set size, we use the rotated interval norm~\cite[Sec.~B.2]{luetzow2025zonoconformal}, which sums the half-widths of the interval hulls over several rotations and thus measures its extent in multiple directions.

\begin{definition}[Rotated interval norm]
\label{def:intNorm}
The rotated interval norm of a zonotope $\mathcal{S} = \langle c, G 
\rangle\subset \mathbb{R}^n$ is
\begin{align*}
    \|\mathcal{S}\|_{V_{1:\nv}} = \sum_{i=1}^{\nv}\mathbf{1}^\top |V_iG|\, \mathbf{1},
\end{align*}
where $|V_iG|$ denotes the element-wise absolute values of $V_iG$, and $V_i\in \mathbb{R}^{n\times n}$, $i=1,\dots,\nv$, are orthogonal matrices. 
\end{definition}\vspace{0.2cm} 
If a deterministic measure is preferred, setting $\nv=1$ with $V_1=\mathbf{I}$ recovers the plain interval norm~\cite{althoff2023conf}, which evaluates the set extension along the coordinate axes only.

\subsection{Problem Statement}

The objective is to identify a model $\sysM$ of the target system $\sysT$ from data such that a prescribed conformance relation $\sysT \preceq_{\mathsf{R}} \sysM$ holds, with $\mathsf{R} \in \{\mathrm{reach},\mathrm{trace}, \mathrm{sim}\}$ as defined in \cref{def:reachConf,def:traceConf,def:simConf}. Among all conformant models, we seek the least conservative one. We quantify conservatism by a function $\mathrm{conservatism}(\sysM)$ that measures the size of the reachable output sets of $\sysM$ (detailed in \cref{sec:linearOpt}), since formal verification often operates on these sets. For simulation conformance, which also transfers state-space properties, a term for the reachable state sets can be added analogously. This yields the optimization problem
\begin{subequations}\label{eq:conf_opt}
\begin{align}
\argmin_{\sysM} ~&\mathrm{conservatism}(\sysM) \label{eq:conf_cost}\\
&\text{s.t. } \text{$\sysT\preceq_{\mathsf{R}}\sysM$}.\label{eq:conf_constr}
\end{align}
\end{subequations}
We assume $\sysT$ is only accessible through the measurement data $\M$, consisting of $\nm$ test cases with the respective output traces $y_{0:\nk-1}^{(m)}(\sysT)$ and, for $\mathsf{R}=\mathrm{sim}$, the state traces $x_{0:\nk}^{(m)}(\sysT)$, $m=1,\dots,\nm$. 
To this end, 
we derive data-driven counterparts of each conformance relation.
If the measurements are noisy, all guarantees refer to the measured quantities rather than to the true states and outputs of $\sysT$.
Since solving~\eqref{eq:conf_opt} exactly is generally intractable, even with the data-driven notions, we derive computationally efficient approximations for the following class of models.
 
\begin{definition}[System with zonotopic uncertainties]\label{def:zds}
    A discrete-time system with zonotopic uncertainties has the form \hspace{-0.5cm}
    \begin{subequations} \label{eq:sysNL}
    \begin{align}    
        x_{k+1} &= f(x_k,u_k) + e_{\mathrm{x},k}, \label{eq:sysNL_x}\\
        y_k &= g(x_k,u_k) + e_{\mathrm{y},k}, \label{eq:sysNL_y}
    \end{align}
    \end{subequations}
    where the functions $f$ and $g$ are at least twice continuously differentiable, $u_k$ consists of control inputs and disturbances, and $e_{\mathrm{x},k}$ and $e_{\mathrm{y},k}$ are additive model uncertainties. Nondeterminism enters through deviations of the initial state $x_0$ and the inputs $u_{0:k}$ from their nominal values $\bar{x}_0^{(m)}$ and $\bar{u}_{0:k}^{(m)}$ and through nonzero model uncertainties, which are bounded for each test case $m$ as
    \begin{subequations}\label{eq:uncSets}
    \begin{align}
        x_0 &\in \bar{x}_0^{(m)} \oplus \XO^{(m)}
    , \\
        u_{0:k} &\in \bar{u}_{0:k}^{(m)} \oplus \mathcal{U}_{\Delta,0:k}^{(m)}
    , \label{eq:uncSets_U}\\
        e_{\mathrm{x},0:k} &\in \mathcal{E}_{\mathrm{x},0:k}^{(m)}, \label{eq:uncSets_ex}
    \\
        e_{\mathrm{y},0:k} &\in \mathcal{E}_{\mathrm{y},0:k}^{(m)},\label{eq:uncSets_ey}
    \end{align}
    \end{subequations}
    where the initial state and input uncertainty sets are zonotopes of the form
    \begin{subequations}\label{eq:uncSetsStructure}
    \begin{align}
    \XO^{(m)} &= \langle \mathbf{0},\, \Gx\mathrm{diag}(\ax)\rangle, \\
    \mathcal{U}_{\Delta,k}^{(m)} &= \langle \mathbf{0},\,
    \Gu\mathrm{diag}(\au)\rangle,
    \end{align}
    \end{subequations}
    with generator templates $\Gx$ and $\Gu$ and scaling factors $\ax\in\mathbb{R}_{\geq 0}^{\nux}$ and $\au\in\mathbb{R}_{\geq 0}^{\nuu}$.
\end{definition}\vspace{0.2cm}
 
We assume $f$ and $g$ to be known a priori, e.g., from first principles or classical system identification, and the generator templates to be predefined, e.g., $\Gx=\mathbf{I}$ and $\Gu=\mathbf{I}$. The model uncertainty sets $\mathcal{E}_{\mathrm{x},0:k}^{(m)}$ and $\mathcal{E}_{\mathrm{y},0:k}^{(m)}$ are derived from $\ax$ and $\au$ to compensate linearization errors, so that only the scaling factors $\alpha := [\ax^\top~\au^\top]^\top$ remain to be identified. Fixing $f$, $g$, $\Gx$, and $\Gu$ is a modeling choice for tractability: we seek the least conservative conformant model within the family parametrized by the scaling factors, although less conservative conformant models may exist outside of it.
 
\begin{examplebox}
\exlabel{Model uncertainties}{ex:pendUnc} The pendulum can be modeled as in \cref{def:zds} by adding $e_{\mathrm{x},k}$ and $e_{\mathrm{y},k}$ to the state and the output equations, respectively. With $\Gx=\mathbf{I}_2$ and $\Gu=\mathbf{I}_3$, we have $\nux=2$ and $\nuu=3$, so that the unknowns are $\alpha\in\mathbb{R}^5_{\geq 0}$. They are the half-widths of the admissible deviations of the initial angle and velocity from $\bar{x}_0$ ($\ax$) and of the normalized torque, the disturbance, and the noise ($\au=[\alpha_\tau~\alpha_w~\alpha_v]^\top$). The bounds $\mathcal{E}_{\mathrm{x},k}^{(m)}$ and $\mathcal{E}_{\mathrm{y},k}^{(m)}$ are not additional unknowns but constructed based on $\alpha$, as shown in \cref{sec:linSets}.
\end{examplebox}

\section{Identification Framework} \label{sec:framework}
 
Solving the identification problem \cref{eq:conf_opt} for any of the three conformance relations requires overcoming the same three obstacles: a) the relations of \cref{def:reachConf,def:traceConf,def:simConf} refer to all test cases in $\mathbb{M}$, whereas only the test cases in the finite set $\M$ are measured; b)~the reachable sets and trace sets of nonlinear models cannot be computed exactly; and c) the optimization over models must remain tractable. We address them by three consecutive steps that apply to all three relations alike: in \cref{sec:confData}, we replace the relations by data-driven counterparts that involve only the measured traces; in \cref{sec:linSets}, we derive linearization-induced sets such that containment of the measurements in them implies conformance of the nonlinear model; and in \cref{sec:linearOpt}, we turn this containment into a linear program in the scaling factors $\alpha$. From here on, we omit the superscript $(m)$ wherever a quantity refers to a single test case $m$, e.g., $\bar{x}_0=\bar{x}_0^{(m)}$,
and sets without a system argument refer to $\sysM$, e.g., $\Y_k=\Y_k(\sysM)$, whereas quantities of $\sysT$ keep the argument $(\sysT)$.
 
\subsection{Data-Driven Conformance Relations} \label{sec:confData}
 
The reachable sets and trace sets of $\sysT$ are generally unknown, and $\mathbb{M}$ is infinite. We therefore replace the relations in \cref{def:reachConf,def:traceConf,def:simConf} by counterparts based on the finite dataset~$\M$. 
 
\begin{definition}[Data-driven reachset conformance]\label{def:reachConfData}
    A system $\sysM$ is a reachset-conformant model of $\sysT$ with respect to the test cases in $\M$, written as $\sysT\preceq_{\mathrm{reach}(\M)}\sysM$, if the reachable output set of $\sysM$ contains every measurement of $\sysT$ for all time steps $k=0,\dots,\nk-1$ and all test cases $m\in\M$:
    \begin{align*}
        \sysT\preceq_{\mathrm{reach}(\M)}\sysM \Leftrightarrow \forall &k=0,\dots,\nk-1,\, m\in\M\colon \\&y_k^{(m)}(\sysT)\in \Y_k^{(m)}.
    \end{align*}
\end{definition}\vspace{0.2cm}
 
\begin{definition}[Data-driven trace conformance {\cite[Def.~6]{kulmburg2026formally}}]\label{def:traceConfData}
    A system $\sysM$ is a trace-conformant model of $\sysT$ with respect to the test cases in $\M$, written as $\sysT \preceq_{\mathrm{trace}(\M)} \sysM$, if the set of output traces of $\sysM$ contains every measured trace of $\sysT$ for all test cases $m \in \M$:
    \begin{align*}
        \sysT \preceq_{\mathrm{trace}(\M)} \sysM \;\Leftrightarrow\;
        \forall\, m \in \M\colon
        y_{0:\nk-1}^{(m)}(\sysT) \in \mathcal{T}_{\mathrm{y}}^{(m)}.
    \end{align*}
\end{definition}\vspace{0.2cm}
Simulation conformance requires measured states and thus excludes initial state uncertainty, i.e., $\ax=\mathbf{0}$, with the nominal initial state set to the measured one, $\bar{x}_0=x_0(\sysT)$. Since the state summarizes the entire past of a system, its future behavior depends only on the current state and the future inputs~\cite{sontag1998mathematical}. A model that attains the measured successor state can therefore be restarted in it, so that it suffices to reproduce every measured pair $z_{k}(\sysT)$. We therefore split the $\nm$ test cases in $\M$ into $\nm\nk$ test cases of horizon one, indexed by $(m,k)$ with $m\in\M$ and $k=0,\dots,\nk-1$: test case $(m,k)$ starts in $\bar{x}_{0}^{(m,k)}=x_{k}^{(m)}(\sysT)$ and applies $\bar{u}_{0}^{(m,k)}=\bar{u}_{k}^{(m)}$. Comprising a single time step, its joint trace set coincides with its joint reachable set~$\Z_{0}^{(m,k)}$.
 
\begin{definition}[Data-driven simulation conformance]\label{def:simConfData}
    A system $\sysM$ is a simulation-conformant model of $\sysT$ with respect to the test cases in $\M$, written as $\sysT\preceq_{\mathrm{sim}(\M)}\sysM$, if
    all measured successor states and outputs are contained in the joint reachable set of $\sysM$:
    \begin{align*}
        \sysT\preceq_{\mathrm{sim}(\M)}\sysM \Leftrightarrow \forall m\in\M,\, k\colon ~ z_{k}^{(m)}(\sysT) \in \Z_{0}^{(m,k)}.
    \end{align*}
\end{definition}\vspace{0.2cm}
 
\cref{def:simConfData} constrains single transitions, whereas \cref{def:simConf} constrains entire joint traces. The two are connected by the following result.
 
\begin{lemma}[Trace reproduction]\label{lem:simTraj}
Let $\sysM$ be a system as in \cref{def:zds} with $\sysT\preceq_{\mathrm{sim}(\M)}\sysM$ and $\bar{x}_0^{(m)}=x_0^{(m)}(\sysT)$ for all $m\in\M$, and let the model uncertainty sets of every test case $m\in\M$ contain those of its one-step test cases, i.e., \begin{align}\label{eq:simReq}
    \mathcal{E}_{\mathrm{x},0}^{(m,k)}\subseteq\mathcal{E}_{\mathrm{x},k}^{(m)} \text{ and } \mathcal{E}_{\mathrm{y},0}^{(m,k)}\subseteq\mathcal{E}_{\mathrm{y},k}^{(m)},~k=0,\dots,\nk-1.
\end{align}
Then $\sysM$ reproduces every measured trace, i.e., $z_{0:\nk-1}^{(m)}(\sysT)\in\mathcal{T}_{\mathrm{z}}^{(m)}$ for all $m\in\M$.
\end{lemma}
\vspace{0.1cm}\textit{Proof:} See Appendix~\ref{app:proofSimTraj}.\vspace{0.2cm}

\begin{examplebox}
\exlabel{Conformance relations}{ex:pendConf} For one measured trace of the pendulum, as shown in \cref{fig:conformanceOverview}, reachset conformance requires every measured angle $y_k$ to lie in the interval $\Y_k$ of that step; trace conformance requires a single output trace
that matches the measured angles at all time steps; and simulation conformance requires, for every $k$, a realization $(\tau_{\Delta,k},w_k,v_k)$ that maps the measured states $(x_{k(1)},x_{k(2)})$ to the measured states and output $(x_{k+1(1)},x_{k+1(2)},y_k)$.
\end{examplebox}

\subsection{Linearization with Error Compensation} \label{sec:linSets}
 
The relations of \cref{def:reachConfData,def:traceConfData,def:simConfData} involve the reachable and trace sets of the nonlinear model \cref{eq:sysNL}, which generally cannot be computed exactly~\cite{Platzer2007}. We therefore impose them on linearization-induced sets, i.e., the reachable and trace sets of the linearization of \cref{eq:sysNL} along the reference trajectory of each test case. The reference trajectory starts at the nominal initial state and follows the nominal inputs with $e_{\mathrm{x},0:\nk-1}=\mathbf{0}$ and $e_{\mathrm{y},0:\nk-1}=\mathbf{0}$,
\begin{align}\label{eq:refTraj}
    \bar{x}_{k+1} = f(\bar{x}_{k},\bar{u}_{k}), \quad \bar{y}_{k}=g(\bar{x}_{k},\bar{u}_{k}).
\end{align}
Adding the negated linearization error to the model uncertainties makes the linearization-induced sets underapproximate those of $\sysM$, so that every measurement they contain is reachable by the nonlinear model.

\begin{theorem}[Joint reachable set]\label{thm:jointReach}
Let $\sysM$ be a system as in \cref{def:zds}. Its joint reachable set for test case $m$ is overapproximated by
\begin{align}\label{eq:ZHat}
    \Z_k \subseteq \widehat{\Z}_k = \bar{\Z}_k \oplus \Z_{\mathrm{e},k} \oplus {\Z}_{\ell,k}
\end{align}
for all $k=0,\dots,\nk-1$, with the linearization-induced joint reachable set $\bar{\Z}_k$, the model uncertainty contribution $\Z_{\mathrm{e},k}$, and the linearization error contribution ${\Z}_{\ell,k}$,
\begin{subequations}\label{eq:ZSets}
\begin{align}
    \bar{\Z}_k &= \{\bar{z}_k\} \oplus \Apropm{k} \XO \oplus \bigoplus_{i=0}^k \Bpropm{k}{i} \mathcal{U}_{\Delta,i}, \label{eq:ZBar}\\
    \Z_{\mathrm{e},k} &= \bigoplus_{i=0}^{k} \Cpropm{k}{i}\mathcal{E}_{\mathrm{x},i} \oplus \Dprop\,\mathcal{E}_{\mathrm{y},k}, \label{eq:ZE}\\
    {\Z}_{\ell,k} &= \bigoplus_{i=0}^{k} \Cpropm{k}{i}\widehat{\mathcal{L}}_{\mathrm{x},i} \oplus \Dprop\,\widehat{\mathcal{L}}_{\mathrm{y},k}, \label{eq:ZL}
\end{align}
\end{subequations}
where $\bar{z}_k=[\bar{x}_{k+1}^{\top}~\bar{y}_{k}^{\top}]^\top$. The construction of the propagation matrices $\Apropm{k}$, $\Bpropm{k}{i}$, $\Cpropm{k}{i}$, and $\Dprop$
is detailed in \cref{eq:propMat} in Appendix~\ref{app:lin}, and $\widehat{\mathcal{L}}_{\mathrm{x},i}$ and $\widehat{\mathcal{L}}_{\mathrm{y},k}$ are enclosures of the attainable linearization errors.
If the model uncertainty sets are chosen as
\begin{align}\label{eq:uncChoice}
    \mathcal{E}_{\mathrm{x},k}\supseteq -\widehat{\mathcal{L}}_{\mathrm{x},k},\quad \mathcal{E}_{\mathrm{y},k}\supseteq -\widehat{\mathcal{L}}_{\mathrm{y},k},
\end{align}
the joint reachable set is underapproximated by the linearization-induced joint reachable set, i.e., 
\begin{align} \label{eq:Zunder}
\Z_k \supseteq \bar{\Z}_k.
\end{align}
\end{theorem}
\vspace{0.1cm}\textit{Proof:} See Appendix~\ref{app:proofJointReach}.\vspace{0.2cm}

The same line of reasoning can be applied to the trace set.
 
\begin{theorem}[Joint trace set]\label{thm:jointTrace}
Let $\sysM$ be a system as in \cref{def:zds} with the uncertainty sets in \cref{eq:uncChoice}. The joint trace set $\mathcal{T}_{\mathrm{z}}$ for test case $m$ is underapproximated by the linearization-induced joint trace set, i.e.,
\begin{align} \label{eq:Tunder}
\mathcal{T}_{\mathrm{z}}\supseteq\bar{\mathcal{T}}_{\mathrm{z}},
\end{align}
with
\begin{subequations}\label{eq:TzBarAll}
\begin{align}
    \bar{\mathcal{T}}_{\mathrm{z}} &= \{\bar{z}_{0:\nk-1}\} \oplus \ATpropm \XO \oplus \BTpropm \mathcal{U}_{\Delta,0:\nk-1}, \label{eq:TzBar}\\
    \ATpropm &= \Apropm{0:\nk-1}, \quad \BTpropm = \begin{bmatrix}
        \Bpropm{0}{0} & & \mathbf{0} \\
        \vdots & \ddots & \\
        \Bpropm{\nk-1}{0} & \cdots & \Bpropm{\nk-1}{\nk-1}
    \end{bmatrix} \label{eq:propMatTrace}
\end{align}
and $\Apropm{k}$, $\Bpropm{k}{i}$ from \cref{eq:propMat} in Appendix~\ref{app:lin}. 
\end{subequations}
\end{theorem}
\vspace{0.1cm}\textit{Proof:} See Appendix~\ref{app:proofJointTrace}.\vspace{0.2cm}

Since the exact linearization error sets are generally intractable, \cref{thm:jointReach,thm:jointTrace} use enclosures $\widehat{\mathcal{L}}_{\mathrm{x},k}$ and $\widehat{\mathcal{L}}_{\mathrm{y},k}$ of them. Any enclosure of the errors attainable on the reachable sets of $\sysM$ is admissible for $\mathsf{R}\in\{\mathrm{reach},\mathrm{trace}\}$, where tighter ones yield less conservative models. For $\mathsf{R}=\mathrm{sim}$, the resulting uncertainty sets must additionally satisfy \cref{eq:simReq}, which holds for the enclosures in \cref{eq:LHat}  in Appendix~\ref{app:lagrange} with $\kappa=1$, see \cref{lem:simReq}.
\begin{remark}[Linear models]\label{rem:linear}
For linear models, $\widehat{\mathcal{L}}_{\mathrm{x},k} = \{\mathbf{0}\}$ and $\widehat{\mathcal{L}}_{\mathrm{y},k} = \{\mathbf{0}\}$, so that
\cref{thm:jointReach,thm:jointTrace} hold with equality.
\end{remark}

Based on these results, the data-driven conformance conditions of \cref{def:reachConfData,def:traceConfData,def:simConfData} can be enforced by requiring containment of the measured data in the linearization-induced sets.
\begin{corollary}[Data-driven conformance]\label{cor:containment}
Let $\sysM$ be a system as in \cref{def:zds} with the uncertainty sets defined in \cref{eq:uncChoice}. Then $\sysT\preceq_{\mathsf{R}(\M)}\sysM$ holds if, for all $m\in\M$ and $k=0,\dots,\nk-1$,\hspace{-0.1cm}
\begin{subequations}\label{eq:containment}
\begin{align}
    y_{k}^{(m)}(\sysT) &\in \bar{\Y}_{k}^{(m)} && \text{for } \mathsf{R}=\mathrm{reach}, \label{eq:reachContainment}\\
    y_{0:\nk-1}^{(m)}(\sysT) &\in \bar{\mathcal{T}}_{\mathrm{y}}^{(m)} && \text{for } \mathsf{R}=\mathrm{trace}, \label{eq:traceContainment}\\
    z_{k}^{(m)}(\sysT) &\in  \bar{\Z}_{0}^{(m,k)} && \text{for } \mathsf{R}=\mathrm{sim}, \label{eq:simContainment}
\end{align}
\end{subequations}
with the linearization-induced sets
\begin{subequations}\label{eq:linObjects}
\begin{align}
    \bar{\Y}_{k} &= \Py\bar{\Z}_k, \label{eq:YBar}\\
    \bar{\mathcal{T}}_{\mathrm{y}} &= \PT\bar{\mathcal{T}}_{\mathrm{z}}, \label{eq:TyBar}\\
    \bar{\Z}_{0}^{(m,k)} &= \bigl\{\bar{z}_{0}^{(m,k)}\bigr\} \oplus \Bprop{0}{0}{m,k}\mathcal{U}_{\Delta,k}, \label{eq:ZBarSim}
\end{align}
\end{subequations}
using the reference vector $\bar{z}_{0}^{(m,k)}$ and the propagation matrix $\Bprop{0}{0}{m,k}$ from \cref{eq:propMat} in Appendix~\ref{app:lin}, for the one-step test case $(m,k)$, and the projection matrices
\begin{align}\label{eq:proj}
    \Py = \bigl[\,\mathbf{0}_{\ny\times\nx}~~\mathbf{I}_{\ny}\,\bigr], \quad
    \PT &= \mathrm{diag}(\Py,\dots,\Py).
\end{align}
\end{corollary}
\vspace{0.1cm}\textit{Proof:} See Appendix~\ref{app:proofContainment}.\vspace{0.2cm}

\begin{figure}[t]
\centering
\input{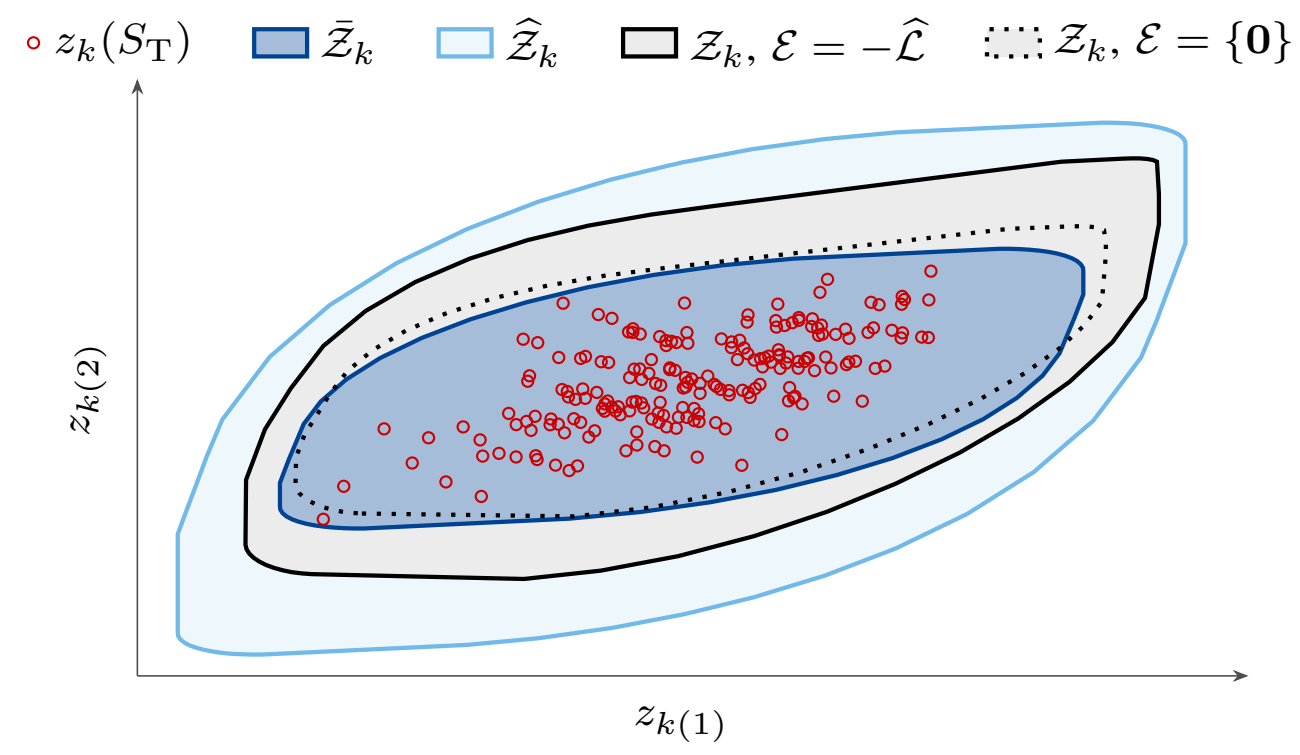}
\caption{Joint reachable sets at time step $k=11$ of a pendulum test case that is measured 200 times (\cref{ex:pendComp}), projected onto their first two dimensions, the states. Compensating the linearization error ($\mathcal{E}=-\widehat{\mathcal{L}}$) enlarges the reachable set of the nonlinear model so that it contains the linearization-induced set $\bar{\Z}_k$ and thus all measurements. The reachable sets ${\Z}_k$ are approximated by simulation.}
\label{fig:errorComp}
\end{figure}

\begin{examplebox}
\exlabel{Error compensation}{ex:pendComp} \Cref{fig:errorComp} illustrates \cref{thm:jointReach,cor:containment} for the pendulum, where all $\nm=200$ visualized test cases share the nominal initial state and inputs. The scaling factors are chosen such that the linearization-induced reachable set $\bar{\Z}_k$ contains all measurements $z_k(\sysT)$. The linearized model could then replace the nonlinear model in subsequent applications, such as verification. To retain the nonlinear model instead, its reachable set must contain $\bar{\Z}_k$. Without additional model uncertainties ($\mathcal{E}=\{\mathbf{0}\}$), the reachable set of the nonlinear model does not overapproximate $\bar{\Z}_k$, so that not all measurements might be reachable. Adding the negated linearization error to the model uncertainties ($\mathcal{E}=-\widehat{\mathcal{L}}$) enlarges this set, so that it overapproximates $\bar{\Z}_k$ by \cref{eq:Zunder} and contains all measurements, as stated by \cref{cor:containment}. An overapproximation of the enlarged set is given by $\widehat{\Z}_k$ according to \cref{eq:ZHat}.
\end{examplebox}

\subsection{Linear Optimization}\label{sec:linearOpt}
 
By \cref{cor:containment}, data-driven conformance is established by the containment of the measured data in the linearization-induced sets as formalized in \cref{eq:containment}. It remains to express this containment as constraints on the unknown parameter $\alpha$ and to quantify the conservatism of a model by a cost function. Both admit a linear formulation, so that the identification problem in \cref{eq:conf_opt} reduces to a linear program.
 
\paragraph{Containment Constraints} \label{sec:constraints}
 
All three containment conditions in \cref{eq:containment} can be expressed as linear constraints in $\alpha$ and a new optimization variable $\beta$, by applying the parametric structure \cref{eq:uncSetsStructure} followed by the zonotope membership condition~\cref{eq:cont_gen}.
 
\begin{proposition}[Containment constraints]\label{prop:constrAll}
For a system $\sysM$ as in \cref{def:zds} with $\ax=\mathbf{0}$ if $\mathsf{R}=\mathrm{sim}$, the containment \cref{eq:containment} holds for a test case $m$ and all $k=0,\dots,\nk-1$ if and only if there exist auxiliary variables $\beta$ such that
\begin{align}\label{eq:constrAll}
    \alpha \geq \mathbf{0}, \qquad |\beta| \leq H_{\mathsf{R}}\alpha, \qquad
    d_{\mathsf{R}} = M_{\mathsf{R}}\beta, 
\end{align}
with the measured deviations $d_{\mathsf{R}}$, the dimension of $\beta$, and the constraint matrices $H_{\mathsf{R}}$ and $M_{\mathsf{R}}$ of \cref{tab:constrMatrices}, which are built from the blocks
\begin{subequations}\label{eq:Q}
\begin{align}
Q_{\mathrm{\alpha},k} &=\begin{bmatrix}
    \mathbf{I}_{\nux} & \mathbf{0}_{\nux \times \nuu} \\
    \mathbf{0}_{(k+1)\nuu \times \nux} & \mathrm{vert}_{k+1}\left( \mathbf{I}_{\nuu}\right)
\end{bmatrix},\\
Q_{\mathrm{\beta},k} &= \begin{bmatrix}
            \Py\Apropm{k} \Gx & \Py\Bpropm{k}{0} \Gu&\cdots&\Py\Bpropm{k}{k} \Gu
        \end{bmatrix},
\end{align}
with $\Apropm{k}$, $\Bpropm{k}{i}$ from \cref{eq:propMat} in Appendix~\ref{app:lin}.
\end{subequations}
\end{proposition}
\vspace{0.1cm}\textit{Proof:} See Appendix~\ref{app:proofConstrAll}.\vspace{0.2cm}

\begin{table*}[t]
\centering
\caption{Variables for \cref{prop:constrAll}, with $y_{\Delta,k}=y_k(\sysT)-\bar{y}_k$, $z_{\Delta,k}=z_k(\sysT)-\bar{z}_0^{(m,k)}$, and  $\beta\in \mathbb{R}^{\dim(\beta)}$.}
\label{tab:constrMatrices}
\small
\begin{tabular}{@{}lllll@{}}
\toprule
$\mathsf{R}$ & $d_{\mathsf{R}}$ & $H_{\mathsf{R}}$ & $M_{\mathsf{R}}$ & $\dim(\beta)$\\
\midrule
$\mathrm{reach}$ & $y_{\Delta,0:\nk-1}$ & $\bigl[Q_{\mathrm{\alpha},0}^\top~\dots~Q_{\mathrm{\alpha},\nk-1}^\top\bigr]^\top$ & $\mathrm{diag}(Q_{\mathrm{\beta},0},\dots,Q_{\mathrm{\beta},\nk-1})$ & ${\sum_{k=0}^{\nk-1}(\nux+(k+1)\nuu)}$\\
$\mathrm{trace}$ & $y_{\Delta,0:\nk-1}$ & $Q_{\mathrm{\alpha},\nk-1}$ & $\bigl[Q_{\mathrm{\beta},k}~~\mathbf{0}_{\ny\times(\nk-1-k)\nuu}\bigr]_{k=0:\nk-1}$ & ${\nux+\nk\nuu}$\\
$\mathrm{sim}$ & $z_{\Delta,0:\nk-1}$ & $\mathrm{vert}_{\nk}\bigl([\mathbf{0}~~\mathbf{I}_{\nuu}]\bigr)$ & $\mathrm{diag}\bigl(\Bprop{0}{0}{m,0}\Gu,\dots,\Bprop{0}{0}{m,\nk-1}\Gu\bigr)$ & ${\nk\nuu}$\\
\bottomrule
\end{tabular}
\end{table*}
The constraints for reachset, trace, and simulation conformance differ only in how the auxiliary variables are shared: for $\mathsf{R}=\mathrm{reach}$, every time step has its own block $\beta_k\in\mathbb{R}^{\nux+(k+1)\nuu}$ and the constraints decouple across time; for $\mathsf{R}=\mathrm{trace}$, a single vector $\beta=[\bx^{\top}~\bu{0}^{\top}~\dots~\bu{\nk-1}^{\top}]^\top$ is shared by all time steps, which couples the constraints; for $\mathsf{R}=\mathrm{sim}$, the test cases have horizon one, so each step has its own $\bu{k}\in\mathbb{R}^{\nuu}$, and the initial state uncertainty is fixed to $\XO=\{\mathbf{0}\}$. Consequently, the stricter trace conformance requires fewer auxiliary variables than reachset conformance (\cref{tab:constrMatrices}). The decision variable $\alpha$ is shared by all test cases.
 
\begin{remark}[Feasibility]\label{rem:feasibility}
Since $|\beta|\leq H_{\mathsf{R}}\alpha$ holds for any $\beta$ once $\alpha$ is large enough, \cref{eq:constrAll} is feasible for any measured data if and only if $M_{\mathsf{R}}$ has full row rank.
As $M_{\mathsf{R}}$ is built from the generator templates and the Jacobians of $f$ and $g$, infeasibility indicates that the templates must be enriched or that $f$ and $g$ in \cref{eq:sysNL} are too restrictive.
\end{remark}\vspace{0.2cm}

\begin{examplebox}
\exlabel{Feasibility}{ex:pendFeas} For the pendulum, $M_{\mathsf{R}}$ has full row rank for all three relations: for $\mathsf{R}\in\{\mathrm{reach},\mathrm{trace}\}$, the noise $v_k$ enters every output directly, and for $\mathsf{R}=\mathrm{sim}$, $\Bprop{0}{0}{m,k}\Gu=\bigl[\begin{smallmatrix}0&1&0\\\Delta t&0&0\\0&0&1\end{smallmatrix}\bigr]$ with $\Gu=\mathbf{I}_3$. For $\Gu=\bigl[\begin{smallmatrix}1&0\\0&0\\0&1\end{smallmatrix}\bigr]$, which forces $w_k=0$, the angle update $x_{k+1(1)}=x_{k(1)}+\Delta t\,x_{k(2)}$ contains no uncertainty and $\Bprop{0}{0}{m,k}\Gu=\bigl[\begin{smallmatrix}0&0\\\Delta t&0\\0&1\end{smallmatrix}\bigr]$ has rank $2<\nx+\ny$, so that the constraints for $\mathsf{R}=\mathrm{sim}$ are infeasible once a measured state pair violates this update.
\end{examplebox}
 
\paragraph{Cost Function} \label{sec:cost} 
 
We seek a conservatism measure in \cref{eq:conf_cost} that is linear in $\alpha$, so that the identification problem can be solved with linear programming. This measure is derived for the reachable output sets; the optional term for the reachable state sets can be derived analogously.
 
By \cref{thm:jointReach}, the reachable output sets of $\sysM$ are enclosed by the projection of \cref{eq:ZHat} onto the output dimensions, $\widehat{\Y}_k = \Py\widehat{\Z}_k = \bar{\Y}_k\oplus\tilde{\Y}_k$, with the linearization-induced set $\bar{\Y}_k=\Py\bar{\Z}_k$ and the remainder $\tilde{\Y}_k=\Py\Z_{\mathrm{e},k}\oplus\Py\Z_{\ell,k}$. The generators of $\bar{\Y}_k$ scale linearly with $\alpha$ by \cref{eq:uncSetsStructure,eq:ZBar}, so its rotated interval norm $\|\bar{\Y}_k\|_{V_{1:\nv}}$ with random orthogonal matrices $V_i\in\mathbb{R}^{\ny\times\ny}$ is linear in $\alpha$. The remainder $\tilde{\Y}_k$, in contrast, contains the linearization error enclosures, which depend nonlinearly on $\alpha$ through the reachable sets (Appendix~\ref{app:lagrange}). We therefore measure conservatism by the size of the linearization-induced set alone,
\begin{align}\label{eq:cost}
    \mathrm{conservatism}(\sysM) = \sum_{m \in \Meval} \sum_{k=0}^{\nk-1} \mu_k \|\bar{\Y}_k^{(m)} \|_{V_{1:\nv}},
\end{align}
with the time-step weights $\mu_k\geq 0$. The evaluation test cases $\Meval$ and the matrices $V_i$ are drawn once and independently of $\M$; since $\Meval$ enters only through its reference trajectories, no measurements for $m \in \Meval$ are needed.
 
\begin{lemma}[Cost function]\label{lem:cost}
Let $\sysM$ be a system as in \cref{def:zds}. The conservatism measure in \cref{eq:cost} is linear in the scaling factors $\alpha\geq\mathbf{0}$,
\begin{align}\label{eq:objective}
    \mathrm{conservatism}(\sysM) = \gamma \alpha,
\end{align}
with the coefficient vector
\begin{align}\label{eq:costCoeff}
    \gamma &= \sum_{m\in\Meval}\sum_{k=0}^{\nk-1} \mu_k \sum_{i=1}^{\nv} \mathbf{1}^\top \notag \\
    &\qquad \begin{bmatrix}
         |V_i\Py\Aprop{k}{m} \Gx| & \sum_{j=0}^{k}|V_i\Py\Bprop{k}{j}{m}\Gu |
    \end{bmatrix},
\end{align}
and $\Apropm{k}$, $\Bpropm{k}{i}$ from \cref{eq:propMat} in Appendix~\ref{app:lin}.
\end{lemma}
\vspace{0.1cm}\textit{Proof:} See Appendix~\ref{app:proofCost}.\vspace{0.2cm}
 
Each entry of $\gamma$ measures how strongly a unit deviation along the corresponding generator enlarges the linearization-induced output sets, weighted by $\mu_k$ and summed over the evaluation test cases and time steps, so that minimizing $\gamma\alpha$ penalizes most the generators to which the outputs are most sensitive. Since $\gamma\geq\mathbf{0}$ by \cref{eq:costCoeff}, minimizing \cref{eq:objective} drives the scaling factors towards zero, so that the identified uncertainty sets are as small as the containment constraints permit. This also justifies neglecting $\tilde{\Y}_k$: since interval arithmetic is inclusion-monotone~\cite[Sec.~4.3]{moore2009introduction}, smaller scaling factors yield smaller enclosures $\widehat{\mathcal{L}}$ (see Appendix~\ref{app:lagrange}) and, hence, a smaller $\tilde{\Y}_k$. The omission does not affect the conformance guarantees either, which are certified by the containment constraints \cref{eq:containment} alone; an omitted cost term can render a model suboptimal but never nonconformant.

\paragraph{Unified Identification Theorem}
 
Combining the cost function of \cref{lem:cost} with the containment constraints of \cref{prop:constrAll} yields a linear program for all three conformance relations.
 
\begin{theorem}[Conformant identification]\label{thm:confId}
Let $\mathsf{R}\in\{\mathrm{reach},\mathrm{trace},\mathrm{sim}\}$ and let $\sysM$ be a system as in \cref{def:zds} with $\XO=\{\mathbf{0}\}$, i.e., $\ax=\mathbf{0}$, if $\mathsf{R}=\mathrm{sim}$. Let the scaling factors be identified by solving the linear program
\begin{subequations}\label{eq:confOpt}
\begin{align}
\alpha^\ast(\M),~\beta^\ast(\M) &= \argmin_{\alpha,\beta} ~ \gamma \alpha \label{eq:confCost}\\
\text{s.t. } &\forall m\in\M\colon ~ \text{\cref{eq:constrAll} for relation } \mathsf{R},\label{eq:confConstr}
\end{align}
\end{subequations}
with $\gamma$ from \cref{eq:costCoeff}, where $\beta$ collects the auxiliary optimization variables $\beta^{(m)}$, for all $m\in\M$. Let the model uncertainty sets be chosen according to \cref{eq:uncChoice}. If \cref{eq:confOpt} is feasible, any of its optimizers $\alpha^\ast(\M)$ yields $\sysT\preceq_{\mathsf{R}(\M)}\sysM$.
\end{theorem}
\vspace{0.1cm}\textit{Proof:} See Appendix~\ref{app:proofConfId}.\vspace{0.2cm}

The resulting identification workflow is summarized in \cref{alg:identification}; the target relation $\mathsf{R}$ only selects the containment constraints of the linear program and, for $\mathsf{R}=\mathrm{sim}$, the splitting into one-step test cases.
 
\begin{algorithm}[t]
\caption{Conformant identification}\label{alg:identification}
\begin{algorithmic}[1]
\REQUIRE $f,g$;\; $\Gx,\Gu$;\; $\mathsf{R}$;\; $\M$;\; $\Meval$
\ENSURE $\sysM$ with $\sysT\preceq_{\mathsf{R}(\M)}\sysM$
\IF{$\mathsf{R}=\mathrm{sim}$}
\STATE $\M\gets\{(m,k)\mid m\in\M,\;k<\nk\}$ \COMMENT{one-step test cases}
\ENDIF
\FORALL{$m\in\M\cup\Meval$}
\STATE $\bar{x}_{0:\nk}^{(m)},\bar{y}_{0:\nk-1}^{(m)}\gets$ \cref{eq:refTraj} \COMMENT{compute reference trajectory}
\STATE $\Aprop{k}{m},\Bprop{k}{i}{m},\Cprop{k}{i}{m},\Dprop\gets$ \cref{eq:propMat} \COMMENT{linearize}
\ENDFOR
\STATE $\gamma\gets$ \cref{eq:costCoeff} \COMMENT{compute cost vector}
\STATE $\alpha^\ast\gets$ \cref{eq:confOpt} \COMMENT{solve linear program}
\RETURN $\sysM$ with $\alpha^\ast$ in \cref{eq:uncSetsStructure}, $\mathcal{E}$ from \cref{eq:uncChoice}
\end{algorithmic}
\end{algorithm}

\begin{remark}[Ordering of the relations]\label{rem:ordering}
Since $\bar{\mathcal{T}}_{\mathrm{y}}\subseteq\bar{\Y}_{0:\nk-1}$, the trace constraints are stricter than the reachset ones, so that the feasible set of the trace program is contained in that of the reachset program and its optimal cost is at least as large. A model identified for $\mathsf{R}=\mathrm{sim}$ is also trace- and reachset-conformant by \cref{lem:simTraj}, yet the feasible set of the simulation program need not be contained in those of the trace and reachset programs, since its containment constraints linearize around measured instead of predicted states.
\end{remark}

\section{Generalization Guarantees} \label{sec:guarantees}
 
The exact relations of \cref{def:reachConf,def:traceConf,def:simConf} require reachable sets and trace sets for all test cases in $\mathbb{M}$ and cannot be certified from finite data. 
Instead, \cref{thm:confId} returns models that are conformant with respect to the measured traces for the test cases in $\M$ and we quantify how well the identified models generalize to unseen test cases. To this end, we use scenario theory, which bounds the probability that the optimizer of a convex program built from $\nm$ test cases violates the constraints imposed by a new, independent test case, i.e., that the identified model is not conformant for it. 
\Cref{subsec:scenario} shows that, although \cref{eq:confOpt} carries $\mathcal{O}(\nm)$ auxiliary variables, the scenario bounds depend only on the number of scaling factors, and \cref{subsec:bounds} derives the resulting coverage bounds. Since \cref{eq:confOpt} has the same structure for all three conformance relations, we treat them jointly.
 
\subsection{The Identification Problems as Scenario Programs} \label{subsec:scenario}
 
The bounds of scenario theory depend on the number of decision variables, which in \cref{eq:confOpt} grows with $\nm$ due to the auxiliary variables $\beta$. To show that the bounds depend only on the dimension of $\alpha$, we derive an equivalent program in $\alpha$ alone. Throughout this section, $\mathcal{A}_0 = \{\alpha \mid \alpha \geq \mathbf{0}\}$ denotes the domain of the scaling factors. For $\mathsf{R}\in\{\mathrm{reach},\mathrm{trace},\mathrm{sim}\}$, $\beta^{(m)}$ are the auxiliary variables of test case $m$ in \cref{prop:constrAll}, and we define the projected constraint set
\begin{align*}
    \mathcal{A}_{\mathsf{R}}^{(m)} = \bigl\{\alpha\in\mathcal{A}_0 \,\big|\, \exists\,\beta^{(m)}\colon \text{\cref{eq:confConstr} holds for test case } m\bigr\},
\end{align*}
which is the set of all scaling factors under which test case $m$ is conformant.
For $\mathsf{R}=\mathrm{sim}$, the constraints of test case $m$ are those of all $\nk$ one-step test cases originating from the measured trace $m$.
Using $\mathcal{A}_{\mathsf{R}}^{(m)}$, the auxiliary variables can be discarded without changing the identification problem:
 
\begin{lemma}[Projected scenario program]\label{lem:projection}
The identification problem in \cref{eq:confOpt} is equivalent to the convex program
\begin{align}\label{eq:projProgram}
    \alpha^\ast(\M) = \argmin_{\alpha\in\mathcal{A}_0} ~ \gamma\alpha \quad \text{s.t. } ~\alpha\in\bigcap_{m\in\M}\mathcal{A}_{\mathsf{R}}^{(m)},
\end{align}
in the sense that both problems have the same optimal value and the same set of optimal $\alpha$. The set $\mathcal{A}_{\mathsf{R}}^{(m)}$ is a closed convex polyhedron that depends on the data of test case $m$ only. 
\end{lemma}
\vspace{0.1cm}\textit{Proof:} See Appendix~\ref{app:proofProjection}.\vspace{0.2cm}
 
This program is never solved in practice, since computing $\mathcal{A}_{\mathsf{R}}^{(m)}$ is expensive, but it allows us to apply scenario theory with $\ntheta=\dim(\alpha)$ decision variables, independently of $\nm$.
 
\subsection{Coverage Bounds} \label{subsec:bounds}
 
Under the following assumptions, scenario theory~\cite{calafiore2006scenario, garatti2022risk} lower-bounds the coverage, i.e., the probability that the identified model is conformant for a new test case. To this end, we define the set of support test cases $\mathrm{supp}(\M) \subseteq \M$ as the set of test cases whose removal would change the optimizer $\alpha^\ast(\M)$ of~\cref{eq:confOpt}.
 
\begin{assumption}\label{ass:independence}
    The test cases in $\M$ and the new test case $\tilde{m}$, each together with its measured trace, are i.i.d.\ samples, and the evaluation set $\Meval$ is independent of $\M$ and $\tilde{m}$.
\end{assumption}\vspace{0.2cm}
 
\begin{assumption}\label{ass:nondegen}
    The scenario program \cref{eq:confOpt} is feasible with a unique optimizer $\alpha^\ast(\M)$. Furthermore, it is non-degenerate with probability one, i.e., solving it only with the support test cases $\mathrm{supp}(\M)$ yields the same solution as using all test cases in $\M$.
\end{assumption}\vspace{0.2cm}
 
The i.i.d.\ requirement demands that the identification data are sampled under the same measure $\mathbb{P}$ under which the model is later used. If the test cases are instead designed, e.g., to excite the system particularly well, the bounds below still hold, but with respect to the distribution induced by the design procedure rather than the operational one. 
Uniqueness is required only for $\alpha^\ast(\M)$, not for the auxiliary variables $\beta$, and can be obtained by a tie-breaking rule~\cite{calafiore2006scenario}. The non-degeneracy condition is mild as it requires that the constraints do not have accumulation points~\cite{romao2023exactFeasibility}. 
 
Probabilities over the data set $\M$ refer to the product measure induced by $\mathbb{P}$, which we denote by $\mathbb{P}$ as well.
 
\begin{theorem}[Conformance guarantees]\label{thm:guarantees}
Let \cref{ass:independence,ass:nondegen} hold. The coverage probability
\begin{align*}
    \eta(\M) = \mathbb{P}\bigl\{\sysT\preceq_{\mathsf{R}(\tilde{m})}\sysM \,\big|\, \M\bigr\},\quad \mathsf{R}\in\{\mathrm{reach},\mathrm{trace},\mathrm{sim}\}
\end{align*}
of the identified model $\sysM$ on an unseen test case $\tilde{m}$ satisfies \hspace{-0.2cm}
\begin{subequations}\label{eq:guarantees}
\begin{align}
    \mathbb{E}\{\eta(\M)\} &\geq 1-\frac{\ntheta}{\nm+1},\label{eq:expectedCov}\\
    \mathbb{P}\{\eta(\M)\geq 1-\epsilon\} &\geq 1-\sum_{i=0}^{\ntheta-1}\binom{\nm}{i}\epsilon^i(1-\epsilon)^{\nm-i},\label{eq:varCov}
\end{align}
\end{subequations}
for any $\epsilon\in(0,1)$, with $\ntheta = \nux+\nuu$ for $\mathsf{R}\in\{\mathrm{reach},\mathrm{trace}\}$ and $\ntheta = \nuu$ for $\mathsf{R}=\mathrm{sim}$. Both bounds are informative only for $\nm>\ntheta$.
\end{theorem}
\vspace{0.1cm}\textit{Proof:} See Appendix~\ref{app:proofGuarantees}.\vspace{0.2cm}

The bound \cref{eq:expectedCov} guarantees that, on average over draws of $\M$, at least this fraction of unseen test cases is covered. The bound \cref{eq:varCov} is stronger: for any prescribed violation level $\epsilon$, its right-hand side, which we call the confidence, bounds the probability over draws of $\M$ that the coverage $\eta(\M)$ is at least $1-\epsilon$. Solving \cref{eq:varCov} for $\epsilon$ provides the coverage level guaranteed at a given confidence, and solving it for $\nm$ provides the number of training test cases required for a target coverage at that confidence. Both bounds tighten as $\nm$ grows relative to $\ntheta$. 
\begin{examplebox}
\exlabel{Coverage guarantee}{ex:pendCov} For the pendulum, $\ntheta=5$ for $\mathsf{R}\in\{\mathrm{reach},\mathrm{trace}\}$ and $\ntheta=3$ for $\mathsf{R}=\mathrm{sim}$. With $\nm=500$ measured traces and $\epsilon=0.05$, \cref{eq:varCov} states that an unseen test case is conformant with probability at least $0.95$, with confidence above $1-10^{-6}$; reducing $\epsilon$ to $0.02$ at the same confidence requires about $1{,}160$ traces.
\end{examplebox}

\begin{remark}[Tightness of the Bounds and Alternatives]\label{rem:waitAndJudge}
The bounds \cref{eq:guarantees} are tight only if $|\mathrm{supp}(\M)|=\ntheta$; since $|\mathrm{supp}(\M)|$ is often smaller, tighter bounds can be obtained a posteriori from the wait-and-judge approach \cite{campi2018waitAndJudge,garatti2022risk}. Alternatively, identifying $\alpha$ on one part of $\M$ and rescaling it by a single parameter calibrated on the remaining $\ncal$ test cases via split conformal prediction \cite{angelopoulos2026theoreticalfoundationsCP,osullivan2025bridgingconformalpredictionscenario} can attain the same expected coverage \cref{eq:expectedCov}, but since the calibration rests on only $\ncal$ test cases, its coverage varies more over draws of $\M$~\cite{luetzow2026mcp}, which weakens its confidence bound \cref{eq:varCov}.
\end{remark}\vspace{0.2cm}

\section{Numerical Experiments} \label{sec:results}

We evaluate our framework along four questions: whether the bounds of \cref{thm:guarantees} hold, whether the linearization error compensation is needed, how the identified models compare with alternatives and across the three relations, and how long the identification takes. We present our empirical answers after describing the common setup of our experiments.

\subsection{Common Setup} \label{sec:resSystems}

The framework is implemented in Python\footnote{The code is provided as supplementary material and will be made publicly available upon acceptance.}, and the linear programs are solved by the interior-point method of HiGHS~\cite{huangfu2018parallelizing}. We evaluate the identified models on unseen test cases $\tilde{\M}$ by their empirical coverage $\hat{\eta}(\tilde{\M})$ and their set size $\sigma(\tilde{\M})$. Unless stated otherwise, we report the coverage of the nonlinear model, i.e., the fraction of test cases for which a nonlinear least-squares search finds a realization of \cref{eq:sysNL} that reproduces the measurements under the respective relation up to a maximum absolute error of $10^{-6}$; since the search is local, this fraction is a lower bound on the coverage. For the linearized model, $\hat{\eta}(\tilde{\M})$ is the fraction of test cases satisfying \cref{eq:containment}. The set size measures how much larger the sets are per output dimension than those of a reference model, i.e., the reachset-conformant model of ConfID in \cref{tab:comparison} and the uncompensated model in \cref{fig:compensation}:
\begin{equation*}
\sigma(\tilde{\M})=\frac{1}{|\tilde{\M}|}\sum_{m\in\tilde{\M}}\Biggl(\frac{\sum_{k}\mathrm{volume}\bigl(\mathcal{H}_k^{(m)}\bigr)}{\sum_{k}\mathrm{volume}\bigl(\mathcal{H}_{\mathrm{ref},k}^{(m)}\bigr)}\Biggr)^{1/\ny},
\end{equation*}
where $\mathcal{H}_k^{(m)}$ is the convex hull of the outputs $y_k$ of the model simulated for extreme and random realizations of test case $m$, an approximation of $\Y_k^{(m)}$.

We enclose the linearization errors by \cref{eq:LHat} with $\kappa=1$ and use the cost in \cref{eq:cost} with $\mu_k=1$, $|\Meval|=10$, and $\nv=8$ ($1$ for scalar outputs), and break ties (\cref{ass:nondegen}) by a fixed random perturbation of $\gamma$ of relative size $10^{-4}$. We evaluate our framework on three systems, a pendulum, a planar quadrotor, and an industrial robot with measured data, which are presented subsequently. Unless stated otherwise, the pendulum and the planar quadrotor use $\nk=10$, $\Gx=\mathbf{I}$, and $\Gu=\mathbf{I}$,
whereas the robot uses $\nk=5$ with its own templates. 

\emph{Pendulum.} We use the system \cref{eq:pendulum} in \cref{ex:pendulum} with $\Delta t=0.05\,$s and $l=1\,$m, so that $\nx=2$, $\nin=3$, and $\ny=1$. The target system starts at the nominal initial state and has a $5\,\%$ longer pendulum than modeled; the nominal initial states, nominal inputs and the target system uncertainties are uniformly sampled with $\bar{x}_{0(1)}\in[-0.3,0.3]\,$rad, $\bar{x}_{0(2)}\in[-0.5,0.5]\,$rad/s, $\bar{\tau}_k\in[-1,1]\,$rad/s$^2$, $\tau_{\Delta,k}\in[-0.1,0.1]\,$rad/s$^2$, $w_k\in[-2,2]\cdot10^{-4}\,$rad, and $v_k\in[-5,5]\cdot10^{-3}\,$rad. \Cref{fig:conformanceOverview} shows the models identified for the three relations from a single measured trace of this system.

\emph{Planar quadrotor.} A quadrotor moving in the vertical plane with the position $x_{(1:2)}$, the pitch angle $x_{(3)}$, their rates $x_{(4:5)}$ and $x_{(6)}$, and two rotors whose thrusts are proportional to the squares of their commands $\zeta_{(1)},\zeta_{(2)}$ evolves as
\begin{equation*}
x_{k+1}=x_k+\Delta t\begin{bmatrix}x_{k(4:6)}\\ \tfrac{c}{m}\bigl(\zeta_{k(1)}^2+\zeta_{k(2)}^2\bigr)\bigl[\begin{smallmatrix}-\sin x_{k(3)}\\ \cos x_{k(3)}\end{smallmatrix}\bigr]-\bigl[\begin{smallmatrix}0\\ g\end{smallmatrix}\bigr]\\ \tfrac{c\,r}{J}\bigl(\zeta_{k(1)}^2-\zeta_{k(2)}^2\bigr)\end{bmatrix}+w_k,
\end{equation*}
with output $y_k=x_{k(1:2)}+v_k$ and input $u=[\zeta^\top~w^\top~v^\top]^\top$ comprising the rotor commands, the process disturbances, and the measurement noise ($\nx=6$, $\nin=10$, $\ny=2$). We use $\Delta t=0.05\,$s, $m=1\,$kg, $J=0.05\,$kg\,m$^2$, $r=0.2\,$m, and $c=mg/2$, so that $\zeta_{(1)}=\zeta_{(2)}=1$ is the hover command. The target system starts at the nominal initial state and has a $5\,\%$ larger mass and moment of inertia than modeled; the nominal initial states, nominal inputs and the target system uncertainties are uniformly sampled with $\bar{x}_{0(1)},\bar{x}_{0(2)}\in[-1,1]\,$m, $\bar{x}_{0(3)}\in[-0.2,0.2]\,$rad, $\bar{x}_{0(4)},\bar{x}_{0(5)}\in[-0.5,0.5]\,$m/s, $\bar{x}_{0(6)}\in[-0.5,0.5]\,$rad/s, $\bar{\zeta}_{k(i)}\in[0.9,1.1]$, and, componentwise, $\zeta_{\Delta,k(i)}\in[-0.05,0.05]$, $w_{k(1:2)}\in[-1,1]\cdot10^{-4}\,$m, $w_{k(3)}\in[-1,1]\cdot10^{-4}\,$rad, $w_{k(4:5)}\in[-5,5]\cdot10^{-3}\,$m/s, $w_{k(6)}\in[-5,5]\cdot10^{-3}\,$rad/s, and $v_k\in[-1,1]\cdot10^{-2}\,$m.

\emph{Robot data.} We use measured data from the KUKA KR300 R2500 benchmark~\cite{weigand2023robotbenchmark}. Its joint angles $x_{(1:6)}$ and joint velocities $x_{(7:12)}$ evolve as
\begin{equation*}
x_{k+1}=x_k+\Delta t\begin{bmatrix}x_{k(7:12)}+\tfrac{\Delta t}{2}a_k\\ a_k\end{bmatrix}+w_k,
\end{equation*}
with $a_k=M(x_{k(1:6)})^{-1}\bigl(U\tau_k-C(x_{k(1:6)},x_{k(7:12)})\,x_{k(7:12)}-G(x_{k(1:6)})-\tau_{\mathrm{H}}(x_{k(1:6)})-\tau_{\mathrm{F}}(x_{k(7:12)})\bigr)$, output $y_k=x_{k(1:6)}+v_k$, and input $u=[\tau^\top~w^\top~v^\top]^\top$ comprising the motor torques, the process disturbances, and the measurement noise ($\nx=12$, $\nin=24$, $\ny=6$). Here, $M$, $C$, and $G$ are the rigid-body terms of the benchmark with the parameters of~\cite{retzler2024augmentation}, $U$ couples the motors to the joints, $\tau_{\mathrm{H}}$ is the weight compensator, and $\tau_{\mathrm{F}(i)}=f_{0(i)}+f_{\mathrm{v}(i)}x_{(6+i)}+f_{\mathrm{c}(i)}\tanh(20\,x_{(6+i)})$ is the friction identified on two reserved trajectories. We low-pass filter the recorded angles, velocities, and torques as in the benchmark and downsample them to $\Delta t=0.1\,$s, so that all three relations, and their guarantees, refer to the filtered signals; every test case is a window at a random start in a random run of one of the $34$ trajectories not used for the friction parameter identification.

Identity generator templates would give $\ntheta\geq24$ and thus low coverage guarantees via \cref{thm:guarantees} for small $\nm$, so we use structured ones without initial-state generators ($\nux=0$), assuming accurate measured initial states, and
\begin{equation*}
\Gu^{\mathrm{reach}}=\Gu^{\mathrm{trace}}=\begin{bmatrix}\mathbf{0}_{6\times6}&\mathbf{0}\\\mathbf{0}_{12\times6}&a_x\\\mathbf{I}_6&\mathbf{0}\end{bmatrix},\quad
\Gu^{\mathrm{sim}}=\begin{bmatrix}\mathbf{0}_{6\times12}&a_\tau\\\mathbf{I}_{12}&\mathbf{0}\\\mathbf{0}_{6\times12}&\mathbf{0}\end{bmatrix},
\end{equation*}
whose block rows act on $\tau$, $w$, and $v$, so that $\ntheta=7$ and $13$; $a_x\in\mathbb{R}^{12}$ and $a_\tau\in\mathbb{R}^{6}$ are the normalized standard deviations of the predicted states and of the measured torques on the two reserved trajectories. For $\mathsf{R}\in\{\mathrm{reach},\mathrm{trace}\}$, every joint receives its own noise generator, and torque uncertainty is omitted, since it would enter through $M(x_{(1:6)})^{-1}U$ and make the linearization error recursion diverge. For $\mathsf{R}=\mathrm{sim}$, the output $y_k=x_{k(1:6)}$ is a measured state and needs no noise, and a torque generator is admissible, since each recursion spans a single step from a measured state.

\subsection{Experiments and Results} \label{sec:resResults}
\subsubsection{Coverage Guarantees} \label{sec:resCoverage}

\begin{figure}[t]
\centering
\input{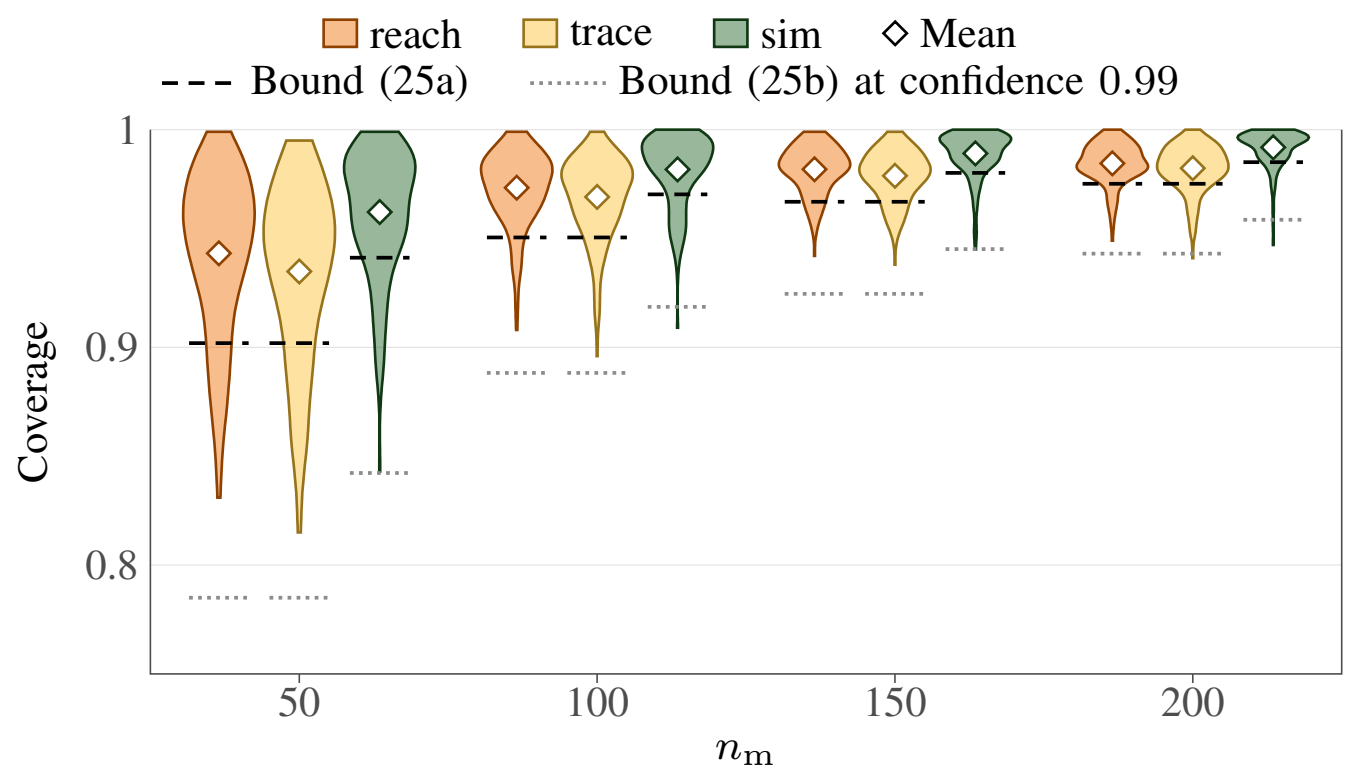}
\vspace{-0.5cm}
\caption{Coverage of the linearized model of the pendulum over $200$ draws per $\nm$.}
\label{fig:coverage}
\vspace{-0.3cm}
\end{figure}

We identify the pendulum for $\nm\in\{50,100,150,200\}$ in $200$ draws of $\M$ each and evaluate the coverage of the linearized model, i.e., the containment \cref{eq:containment} that \cref{thm:guarantees} bounds, on $1000$ unseen test cases. \Cref{fig:coverage} shows that both bounds of \cref{thm:guarantees} hold empirically for all three relations. The mean coverage exceeds \cref{eq:expectedCov} for all relations and all $\nm$, and only $5$ of the $2400$ draws fall below \cref{eq:varCov} at confidence $0.99$, whereas up to $1\,\%$ are permitted. The bounds are conservative because the actual number of support test cases is on average $2.2$ to $2.7$ instead of $\ntheta=5$ for $\mathsf{R}\in\{\mathrm{reach},\mathrm{trace}\}$ and $2.8$ to $2.9$ instead of $\ntheta=3$ for $\mathsf{R}=\mathrm{sim}$ (\cref{rem:waitAndJudge}). For $\mathsf{R}\in\{\mathrm{reach},\mathrm{trace}\}$, some of the optimal scaling factors are zero, e.g., that of the initial angular velocity in all draws; the corresponding generators are unused, which suggests that sparser generator templates would suffice.

\subsubsection{Linearization Error Compensation} \label{sec:resCompensation}

\begin{figure}[t]
\centering
\input{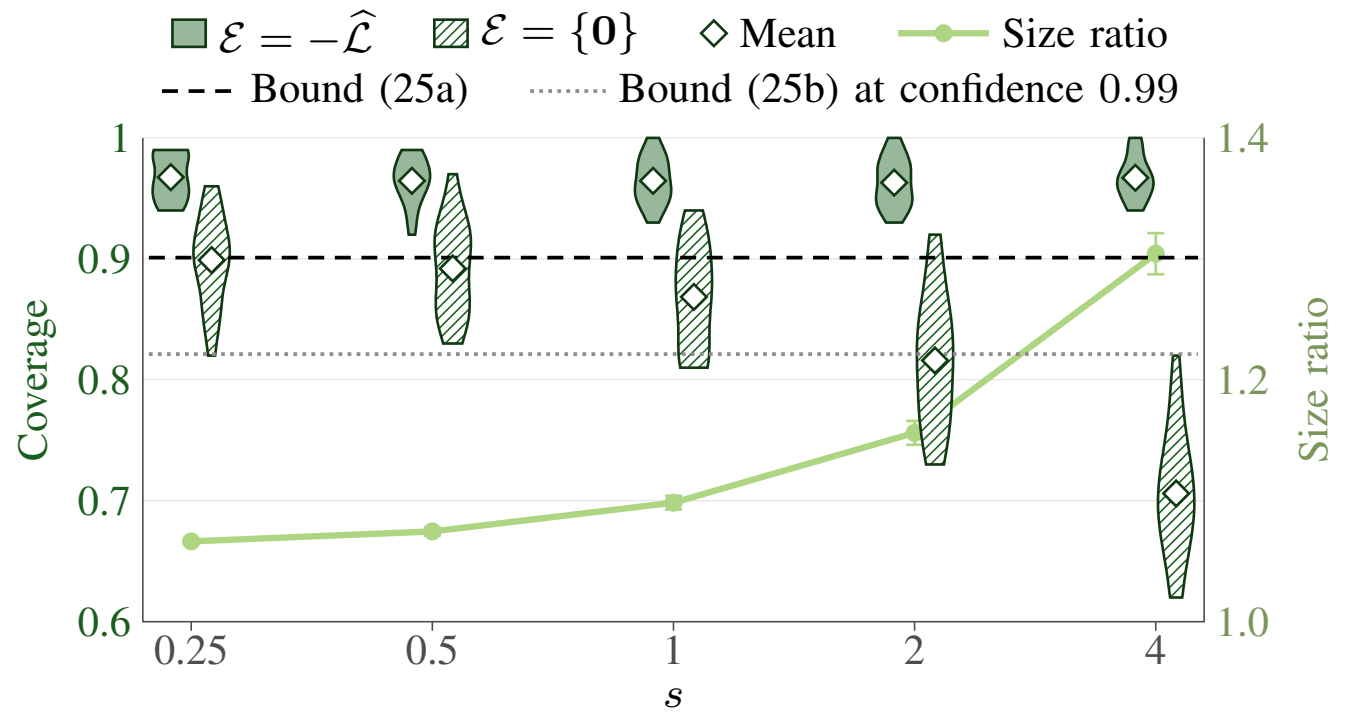}
\vspace{-0.5cm}
\caption{Coverage with and without compensation and ratio of the sizes of both models for the planar quadrotor and $\mathsf{R}=\mathrm{sim}$ over $20$ draws per deviation scale $s$.}
\label{fig:compensation}
\vspace{-0.3cm}
\end{figure}

We scale the command errors, disturbances, and noise of the planar quadrotor target system by $s\in\{0.25,0.5,1,2,4\}$, so that the linearization error grows with $s$, and identify $\alpha^\ast(\M)$ in $20$ draws with $\nm=100$. Since \cref{eq:confOpt} does not depend on $\mathcal{E}$, the compensated ($\mathcal{E}=-\widehat{\mathcal{L}}$) and uncompensated ($\mathcal{E}=\{\mathbf{0}\}$) models use the same $\alpha^\ast(\M)$. As shown in \cref{fig:compensation}, only the compensation guarantees conformance of the nonlinear model.
With compensation, all training test cases are reproduced at every scale, as guaranteed by \cref{cor:containment}, and about $95\,\%$ of $100$ unseen ones for $\mathsf{R}\in\{\mathrm{reach},\mathrm{trace}\}$ and $96\,\%$ for $\mathsf{R}=\mathrm{sim}$. Without compensation, about $95\,\%$ of the training and $93\,\%$ of the unseen test cases are still reproduced for $\mathsf{R}\in\{\mathrm{reach},\mathrm{trace}\}$, since the uncertainties of an entire trace can absorb most of the linearization error, but for $\mathsf{R}=\mathrm{sim}$, the coverage drops from about $90\,\%$ at $s=0.25$ to $70\,\%$ at $s=4$. The price of compensation is larger sets, with a size ratio growing from $1.07$ to $1.30$ as the linearization error enclosure \cref{eq:remainderSet} grows with $s$.

\subsubsection{Comparison} \label{sec:resComparison}

\begin{table*}[t]
\definecolor{tabReach}{RGB}{239,108,0}
\definecolor{tabTrace}{RGB}{251,192,45}
\definecolor{tabSim}{RGB}{27,94,32}
\centering
\caption{Comparison over 50 draws: empirical coverage $\hat{\eta}_\mathsf{R}$ under each relation $\mathsf{R}$, with the bound \cref{eq:expectedCov} on its expectation in parentheses, and set size $\sigma$ relative to our reachset-conformant model (mean $\pm$ standard deviation). Shaded: guaranteed by the targeted relation, otherwise gray; $\geq$: coverage of the linearized model where the error recursion diverges.}
\label{tab:comparison}
\footnotesize
\setlength{\tabcolsep}{3pt}
\begin{tabular}{@{}llrrrrrrrrrrrr@{}}
\toprule
\textbf{Method} & \textbf{Target} & \multicolumn{4}{c}{\textbf{Pendulum}} & \multicolumn{4}{c}{\textbf{Planar quadrotor}} & \multicolumn{4}{c}{\textbf{Robot data}} \\
\cmidrule(lr){3-6} \cmidrule(lr){7-10} \cmidrule(lr){11-14}
 & & \multicolumn{1}{l}{\cellcolor{tabReach!25}$\hat{\eta}_{\mathrm{reach}}$} & \multicolumn{1}{l}{\cellcolor{tabTrace!25}$\hat{\eta}_{\mathrm{trace}}$} & \multicolumn{1}{l}{\cellcolor{tabSim!25}$\hat{\eta}_{\mathrm{sim}}$} & \multicolumn{1}{l}{$\sigma$} & \multicolumn{1}{l}{\cellcolor{tabReach!25}$\hat{\eta}_{\mathrm{reach}}$} & \multicolumn{1}{l}{\cellcolor{tabTrace!25}$\hat{\eta}_{\mathrm{trace}}$} & \multicolumn{1}{l}{\cellcolor{tabSim!25}$\hat{\eta}_{\mathrm{sim}}$} & \multicolumn{1}{l}{$\sigma$} & \multicolumn{1}{l}{\cellcolor{tabReach!25}$\hat{\eta}_{\mathrm{reach}}$} & \multicolumn{1}{l}{\cellcolor{tabTrace!25}$\hat{\eta}_{\mathrm{trace}}$} & \multicolumn{1}{l}{\cellcolor{tabSim!25}$\hat{\eta}_{\mathrm{sim}}$} & \multicolumn{1}{l}{$\sigma$} \\
 & & \multicolumn{1}{l}{\cellcolor{tabReach!25}(0.98)} & \multicolumn{1}{l}{\cellcolor{tabTrace!25}(0.98)} & \multicolumn{1}{l}{\cellcolor{tabSim!25}(0.99)} &  & \multicolumn{1}{l}{\cellcolor{tabReach!25}(0.92)} & \multicolumn{1}{l}{\cellcolor{tabTrace!25}(0.92)} & \multicolumn{1}{l}{\cellcolor{tabSim!25}(0.95)} &  & \multicolumn{1}{l}{\cellcolor{tabReach!25}(0.97)} & \multicolumn{1}{l}{\cellcolor{tabTrace!25}(0.97)} & \multicolumn{1}{l}{\cellcolor{tabSim!25}(0.94)} &  \\
\midrule
\multirow{3}{*}{\emph{ConfID}} & \cellcolor{tabReach!25}reach & \cellcolor{tabReach!25}0.99\,{\scriptsize$\pm$.02} & \textcolor{black!45}{0.00\phantom{\,{\scriptsize$\pm$.00}}} & \textcolor{black!45}{0.00\phantom{\,{\scriptsize$\pm$.00}}} & 1.00\phantom{\,{\scriptsize$\pm$.00}} & \cellcolor{tabReach!25}0.97\,{\scriptsize$\pm$.02} & \textcolor{black!45}{0.00\phantom{\,{\scriptsize$\pm$.00}}} & \textcolor{black!45}{0.00\phantom{\,{\scriptsize$\pm$.00}}} & 1.00\phantom{\,{\scriptsize$\pm$.00}} & \cellcolor{tabReach!25}0.98\,{\scriptsize$\pm$.02} & \textcolor{black!45}{0.98\,{\scriptsize$\pm$.02}} & \textcolor{black!45}{0.00\phantom{\,{\scriptsize$\pm$.00}}} & 1.00\phantom{\,{\scriptsize$\pm$.00}} \\
 & \cellcolor{tabTrace!25}trace & \cellcolor{tabReach!25}0.99\,{\scriptsize$\pm$.02} & \cellcolor{tabTrace!25}0.98\,{\scriptsize$\pm$.02} & \textcolor{black!45}{0.05\,{\scriptsize$\pm$.10}} & 1.00\phantom{\,{\scriptsize$\pm$.00}} & \cellcolor{tabReach!25}0.98\,{\scriptsize$\pm$.02} & \cellcolor{tabTrace!25}0.97\,{\scriptsize$\pm$.02} & \textcolor{black!45}{0.00\phantom{\,{\scriptsize$\pm$.00}}} & 1.01\,{\scriptsize$\pm$.02} & \cellcolor{tabReach!25}0.98\,{\scriptsize$\pm$.02} & \cellcolor{tabTrace!25}0.98\,{\scriptsize$\pm$.02} & \textcolor{black!45}{0.00\phantom{\,{\scriptsize$\pm$.00}}} & 1.00\phantom{\,{\scriptsize$\pm$.00}} \\
 & \cellcolor{tabSim!25}sim & \cellcolor{tabReach!25}1.00\phantom{\,{\scriptsize$\pm$.00}} & \cellcolor{tabTrace!25}1.00\phantom{\,{\scriptsize$\pm$.00}} & \cellcolor{tabSim!25}0.99\,{\scriptsize$\pm$.01} & 1.30\,{\scriptsize$\pm$.06} & \cellcolor{tabReach!25}1.00\,{\scriptsize$\pm$.01} & \cellcolor{tabTrace!25}1.00\,{\scriptsize$\pm$.01} & \cellcolor{tabSim!25}0.98\,{\scriptsize$\pm$.02} & 1.67\,{\scriptsize$\pm$.07} & \cellcolor{tabReach!25}$\geq$0.99\,{\scriptsize$\pm$.01} & \cellcolor{tabTrace!25}$\geq$0.98\,{\scriptsize$\pm$.02} & \cellcolor{tabSim!25}0.97\,{\scriptsize$\pm$.02} & \multicolumn{1}{c}{n/a} \\
\cmidrule{1-14}
\multirow{3}{*}{\begin{tabular}[c]{@{}l@{}}\emph{Ablation}\\\emph{with data}\\\emph{split}\end{tabular}} & \cellcolor{tabReach!25}reach & \cellcolor{tabReach!25}0.97\,{\scriptsize$\pm$.03} & \textcolor{black!45}{0.00\phantom{\,{\scriptsize$\pm$.00}}} & \textcolor{black!45}{0.00\phantom{\,{\scriptsize$\pm$.00}}} & 1.00\,{\scriptsize$\pm$.03} & \cellcolor{tabReach!25}0.93\,{\scriptsize$\pm$.06} & \textcolor{black!45}{0.00\phantom{\,{\scriptsize$\pm$.00}}} & \textcolor{black!45}{0.00\phantom{\,{\scriptsize$\pm$.00}}} & 1.00\,{\scriptsize$\pm$.04} & \cellcolor{tabReach!25}0.97\,{\scriptsize$\pm$.03} & \textcolor{black!45}{0.97\,{\scriptsize$\pm$.03}} & \textcolor{black!45}{0.00\phantom{\,{\scriptsize$\pm$.00}}} & 1.01\,{\scriptsize$\pm$.19} \\
 & \cellcolor{tabTrace!25}trace & \cellcolor{tabReach!25}0.98\,{\scriptsize$\pm$.02} & \cellcolor{tabTrace!25}0.98\,{\scriptsize$\pm$.03} & \textcolor{black!45}{0.05\,{\scriptsize$\pm$.10}} & 1.00\,{\scriptsize$\pm$.03} & \cellcolor{tabReach!25}0.94\,{\scriptsize$\pm$.07} & \cellcolor{tabTrace!25}0.93\,{\scriptsize$\pm$.08} & \textcolor{black!45}{0.00\phantom{\,{\scriptsize$\pm$.00}}} & 1.01\,{\scriptsize$\pm$.04} & \cellcolor{tabReach!25}0.96\,{\scriptsize$\pm$.05} & \cellcolor{tabTrace!25}0.96\,{\scriptsize$\pm$.05} & \textcolor{black!45}{0.00\phantom{\,{\scriptsize$\pm$.00}}} & 1.02\,{\scriptsize$\pm$.19} \\
 & \cellcolor{tabSim!25}sim & \cellcolor{tabReach!25}1.00\phantom{\,{\scriptsize$\pm$.00}} & \cellcolor{tabTrace!25}1.00\phantom{\,{\scriptsize$\pm$.00}} & \cellcolor{tabSim!25}0.99\,{\scriptsize$\pm$.01} & 1.31\,{\scriptsize$\pm$.06} & \cellcolor{tabReach!25}1.00\phantom{\,{\scriptsize$\pm$.00}} & \cellcolor{tabTrace!25}1.00\phantom{\,{\scriptsize$\pm$.00}} & \cellcolor{tabSim!25}0.98\,{\scriptsize$\pm$.02} & 1.68\,{\scriptsize$\pm$.07} & \cellcolor{tabReach!25}$\geq$0.97\,{\scriptsize$\pm$.04} & \cellcolor{tabTrace!25}$\geq$0.96\,{\scriptsize$\pm$.05} & \cellcolor{tabSim!25}0.93\,{\scriptsize$\pm$.07} & \multicolumn{1}{c}{n/a} \\
\cmidrule{1-14}
\multirow{2}{*}{\begin{tabular}[c]{@{}l@{}}\emph{Residual}\\\emph{box}\end{tabular}} & \cellcolor{tabTrace!25}trace & \cellcolor{tabReach!25}0.98\,{\scriptsize$\pm$.02} & \cellcolor{tabTrace!25}0.98\,{\scriptsize$\pm$.02} & \textcolor{black!45}{0.00\phantom{\,{\scriptsize$\pm$.00}}} & 1.61\,{\scriptsize$\pm$.18} & \cellcolor{tabReach!25}0.91\,{\scriptsize$\pm$.09} & \cellcolor{tabTrace!25}0.91\,{\scriptsize$\pm$.09} & \textcolor{black!45}{0.00\phantom{\,{\scriptsize$\pm$.00}}} & 1.30\,{\scriptsize$\pm$.11} & \cellcolor{tabReach!25}0.96\,{\scriptsize$\pm$.03} & \cellcolor{tabTrace!25}0.96\,{\scriptsize$\pm$.03} & \textcolor{black!45}{0.00\phantom{\,{\scriptsize$\pm$.00}}} & 1.10\,{\scriptsize$\pm$.22} \\
 & \cellcolor{tabSim!25}sim & \cellcolor{tabReach!25}1.00\phantom{\,{\scriptsize$\pm$.00}} & \cellcolor{tabTrace!25}1.00\phantom{\,{\scriptsize$\pm$.00}} & \cellcolor{tabSim!25}0.99\,{\scriptsize$\pm$.01} & 1.31\,{\scriptsize$\pm$.07} & \cellcolor{tabReach!25}1.00\phantom{\,{\scriptsize$\pm$.00}} & \cellcolor{tabTrace!25}1.00\phantom{\,{\scriptsize$\pm$.00}} & \cellcolor{tabSim!25}0.97\,{\scriptsize$\pm$.03} & 1.75\,{\scriptsize$\pm$.14} & \cellcolor{tabReach!25}0.97\,{\scriptsize$\pm$.03} & \cellcolor{tabTrace!25}0.96\,{\scriptsize$\pm$.04} & \cellcolor{tabSim!25}0.94\,{\scriptsize$\pm$.06} & 1.26\,{\scriptsize$\pm$.34} \\
\bottomrule
\end{tabular}
\end{table*}

We compare \emph{ConfID} with an ablation and a baseline that keep the nominal model and differ only in how its uncertainty sets are obtained. The \emph{Ablation} solves \cref{eq:confOpt} on $\nm-\ncal$ test cases and rescales $\alpha$ by split conformal calibration on the remaining $\ncal$ test cases. We choose $\ncal=\lceil (\nm+1)/\ntheta\rceil-1$, the smallest number of test cases for which \cref{eq:expectedCov} with $\ntheta=1$ guarantees at least the expected coverage of \emph{ConfID}.
Our baseline, called \emph{Residual box}, transfers the idea of interval enclosures in~\cite{kochdumper2020conf,kochdumper2022conformant} to the given nominal nonlinear model: it uses the interval hull of the prediction errors on $\nm-\ncal$ test cases as $\mathcal{E}$ and rescales it by the same calibration on the remaining $\ncal$ test cases. For $\mathsf{R}\in\{\mathrm{reach},\mathrm{trace}\}$, its model follows the nominal trajectory with output errors bounded independently per step, so that its trace set is the Cartesian product of its reachable sets and both relations yield the same model, listed under trace. We cannot compare against the other methods reviewed in \cref{sec:relWork}, since each replaces the nominal model by a finite-state abstraction, a piecewise affine model, or a linear model, relies on a specific reachability algorithm, or only tests conformance for user-chosen uncertainty sets. 
All methods use the same $\nm=200$ test cases in $50$ draws of $\M$ on all three systems, and we evaluate the identified models on $100$ unseen test cases per draw. 

\Cref{tab:comparison} shows that the reachset-conformant models have sets of nearly the same size as the trace-conformant models but reproduce none of the unseen traces, whereas the simulation-conformant models cover all three relations at the price of larger sets. \emph{ConfID} matches the set sizes of \emph{Ablation} and \emph{Residual box} but covers unseen test cases more reliably. All methods attain the expected coverage \cref{eq:expectedCov} up to about $0.01$, \emph{Ablation} with sets of the same size as those of \emph{ConfID} and \emph{Residual box} with larger ones. Since the confidence bound \cref{eq:varCov} of \emph{ConfID} rests on all $\nm=200$ test cases, it is always larger than the bound of \emph{Ablation} and \emph{Residual box}, which rests only on their $\ncal$ calibration cases. For example, for the reachset program of the quadrotor, \cref{eq:varCov} guarantees $\mathbb{P}\{\eta\geq 0.9\}\geq 0.86$ for \emph{ConfID} instead of $\geq 0.72$ for \emph{Ablation} and \emph{Residual box}; accordingly, \emph{ConfID} attains at least $90\,\%$ coverage in every draw, whereas \emph{Ablation} and \emph{Residual box} fall below it in up to $12$ and $16$ of the $50$ draws, respectively. 

For the robot, a limitation of the error enclosure from Appendix~\ref{app:lagrange} becomes apparent: for the models identified for $\mathsf{R}=\mathrm{sim}$, the linearization error recursion diverges over the multi-step evaluation test cases, since the second derivative of the friction term $\tanh(20\,x_{(6+i)})$ becomes very large near zero velocity. Their reach and trace entries therefore give the coverage of the linearized model, a lower bound on the coverage of the nonlinear model. The recursion also diverges for the reachset- and trace-conformant robot models in one draw of \emph{ConfID} and in up to two of \emph{Ablation}; \cref{tab:comparison} averages over the remaining draws.

\subsubsection{Computation Time} \label{sec:resScalability}

\begin{table}[t]
\definecolor{tabReach}{RGB}{239,108,0}
\definecolor{tabTrace}{RGB}{251,192,45}
\definecolor{tabSim}{RGB}{27,94,32}
\centering
\caption{Identification time in seconds on an Apple M5 Pro over 20 draws (mean $\pm$ standard deviation).}
\label{tab:scalability}
\footnotesize
\setlength{\tabcolsep}{4pt}
\begin{tabular}{@{}lrrrrr@{}}
\toprule
\textbf{System} & $n_\mathrm{k}$ & $n_\mathrm{m}$ & \multicolumn{1}{l}{\cellcolor{tabReach!25}reach} & \multicolumn{1}{l}{\cellcolor{tabTrace!25}trace} & \multicolumn{1}{l}{\cellcolor{tabSim!25}sim} \\
\midrule
\multirow{5}{*}{\textbf{Pendulum}} & 10 & 100 & 0.39\,{\scriptsize$\pm$.03} & 0.09\,{\scriptsize$\pm$.00} & 0.10\,{\scriptsize$\pm$.00} \\
 & 20 & 100 & 1.66\,{\scriptsize$\pm$.08} & 0.28\,{\scriptsize$\pm$.03} & 0.18\,{\scriptsize$\pm$.00} \\
 & 50 & 100 & 15\,{\scriptsize$\pm$1} & 1.38\,{\scriptsize$\pm$.20} & 0.45\,{\scriptsize$\pm$.01} \\
 & 10 & 200 & 0.88\,{\scriptsize$\pm$.04} & 0.19\,{\scriptsize$\pm$.01} & 0.18\,{\scriptsize$\pm$.00} \\
 & 10 & 400 & 2.11\,{\scriptsize$\pm$.13} & 0.50\,{\scriptsize$\pm$.03} & 0.36\,{\scriptsize$\pm$.01} \\
\midrule
\multirow{3}{*}{\textbf{Planar quadrotor}} & 10 & 200 & 6.67\,{\scriptsize$\pm$4.13} & 1.00\,{\scriptsize$\pm$.10} & 0.48\,{\scriptsize$\pm$.09} \\
 & 20 & 200 & 98\,{\scriptsize$\pm$109} & 4.83\,{\scriptsize$\pm$.57} & 1.22\,{\scriptsize$\pm$.20} \\
 & 10 & 400 & 21\,{\scriptsize$\pm$17} & 2.52\,{\scriptsize$\pm$.30} & 1.19\,{\scriptsize$\pm$.27} \\
\midrule
\multirow{3}{*}{\textbf{Robot data}} & 5 & 200 & 1.05\,{\scriptsize$\pm$.01} & 0.96\,{\scriptsize$\pm$.00} & 2.60\,{\scriptsize$\pm$.02} \\
 & 20 & 200 & 8.15\,{\scriptsize$\pm$2.67} & 3.85\,{\scriptsize$\pm$.29} & 11\,{\scriptsize$\pm$0} \\
 & 5 & 400 & 2.29\,{\scriptsize$\pm$.09} & 2.00\,{\scriptsize$\pm$.01} & 5.51\,{\scriptsize$\pm$.06} \\
\bottomrule
\end{tabular}
\end{table}

The identification times are presented in \cref{tab:scalability}. The trace and simulation programs are solved within seconds, whereas the reachset program scales worst with $\nk$, as predicted by its number of auxiliary variables (\cref{sec:linearOpt}). The reachset program thus takes longest for the pendulum and the quadrotor, whereas for the robot the simulation program does, with at most $11\,$s.
 
\section{Conclusion} \label{sec:conclusion}
 
We presented ConfID, a framework that identifies the uncertainty sets of a possibly nonlinear model such that it is reachset-conformant, trace-conformant, or simulation-conformant with respect to measured traces. Unlike existing identification methods, which target a single relation and are mostly confined to linear dynamics, finite abstractions, or a specific reachability algorithm, our framework treats all three relations with the same model class, parametrization, and solver: in each case, the identification is a linear program. In contrast to approaches that linearize the dynamics without accounting for the linearization error, its compensation carries the conformance to the nonlinear model, and scenario theory provides distribution-free bounds on the coverage of unseen test cases. In our experiments, the coverage satisfied both bounds, the compensation preserved conformance of the nonlinear model, and the coverage varied less than with data splitting, with smaller trace sets than a residual-box baseline.

The framework has several limitations that future work should address. First, it returns the least conservative model only within the family defined by the fixed nominal model and the manually chosen templates; since the number of scaling factors enters the bounds directly, small data sets require structured templates. Second, the linearization error compensation inherits the conservatism of the linearization error enclosure. In particular, the enclosures by second-order Lagrange remainders used in this work can diverge over multi-step horizons for steep nonlinearities, such as the friction of the robot, which calls for tighter enclosures. Moreover, standard reachability algorithms applied to the identified model enclose its linearization error once more, although its model uncertainties already compensate for it, so that the sets used for verification are larger than necessary. Third, the guarantees hold only if the identification data are i.i.d.\ samples from the distribution under which the model is later used, and thus not under changed operating conditions. Fourth, simulation conformance requires state measurements and inputs rich enough to reproduce every measured transition. Finally, the number of auxiliary variables of the reachset program grows quadratically with $\nk$, which underapproximative order reduction~\cite{luetzow2025underapprox} could mitigate while preserving the guarantees.

\appendices
\section{Linearization Details} \label{app:lin}
 
This appendix collects the derivation of the propagation matrices of \cref{thm:jointReach}, and of overapproximations $\widehat{\mathcal{L}}_{\mathrm{x},k}\supseteq {\mathcal{L}}_{\mathrm{x},k}$ and $\widehat{\mathcal{L}}_{\mathrm{y},k}\supseteq {\mathcal{L}}_{\mathrm{y},k}$ of the linearization error sets.
 
\subsection{Taylor Expansion and Propagation Matrices}
 
A Taylor expansion with Lagrange remainder \cite[Sec.~3.1]{althoff2013nonlin},~\cite{Berz1998} of $f$ and $g$ around the reference point $(\bar{x}_{k},\bar{u}_{k})$ of \cref{eq:refTraj}, inserted into \cref{eq:sysNL}, leads to
\begin{subequations}\label{eq:taylor}
\begin{align}
    x_{k+1} &= \bar{x}_{k+1} + A_{k} x_{\Delta,k} + B_{k} u_{\Delta,k} + e_{\mathrm{x},k} + \ell_{\mathrm{x},k}, \label{eq:linModel_x}\\
    y_{k} &= \bar{y}_{k} + C_{k} x_{\Delta,k} + D_{k} u_{\Delta,k} + e_{\mathrm{y},k} + \ell_{\mathrm{y},k}, \label{eq:linModel_y}
\end{align}
\end{subequations}
with the deviations $x_{\Delta,k}=x_k-\bar{x}_{k}$ and $u_{\Delta,k}=u_k-\bar{u}_{k}$, the linearization errors $\ell_{\mathrm{x},k}\in{\mathcal{L}}_{\mathrm{x},k}$ and $\ell_{\mathrm{y},k}\in{\mathcal{L}}_{\mathrm{y},k}$, which collect all terms of order two and higher, and the Jacobians
\begin{align*}
    A_{k} = \nabla_{x}f, ~ B_{k} = \nabla_{u}f, ~ C_{k} = \nabla_{x}g, ~D_{k} = \nabla_{u}g,
\end{align*}
evaluated at $(\bar{x}_{k},\bar{u}_{k})$.
Resolving the recursion \cref{eq:linModel_x} and collecting the successor state and the output in $z_k$ give the explicit input--output relation
\begin{align}\label{eq:linSolution}
    z_k &= \bar{z}_k + \Apropm{k} x_{\Delta,0} + \sum_{i=0}^k \Bpropm{k}{i} u_{\Delta,i} \notag \\
    &\quad + \sum_{i=0}^{k} \Cpropm{k}{i} \bigl(e_{\mathrm{x},i} + \ell_{\mathrm{x},i}\bigr) + \Dprop\bigl(e_{\mathrm{y},k} + \ell_{\mathrm{y},k}\bigr),
\end{align}
with the joint reference value $\bar{z}_k=[\bar{x}_{k+1}^{\top}~\bar{y}_{k}^{\top}]^\top$ and the joint propagation matrices
\begin{subequations}\label{eq:propMat}
\begin{align}
    \Apropm{k} &= \begin{cases}
        \bigl[A_{k}^{\top}~C_{k}^{\top}\bigr]^\top \Px\,\Apropm{k-1} & k\geq0, \\[2pt]
        \Px^\top & k=-1,
    \end{cases} \label{eq:propMatE}\\
    \Bpropm{k}{i} &= \begin{cases}
        \bigl[A_{k}^{\top}~C_{k}^{\top}\bigr]^\top \Px\,\Bpropm{k-1}{i} & i<k, \\[2pt]
        \bigl[B_{k}^{\top}~D_{k}^{\top}\bigr]^\top & i=k,
    \end{cases} \label{eq:propMatF}\\
    \Cpropm{k}{i} &= \begin{cases}
        \bigl[A_{k}^{\top}~C_{k}^{\top}\bigr]^\top \Px\,\Cpropm{k-1}{i} & i<k, \\[2pt]
        \Px^\top & i=k,
    \end{cases} \label{eq:propMatG}\\
    \Dprop &= \Py^\top, \label{eq:propMatH}
\end{align}
\end{subequations}
for $0\leq i\leq k$, with the state projection $\Px = \bigl[\,\mathbf{I}_{\nx}~~\mathbf{0}_{\nx\times\ny}\,\bigr]$.

\subsection{Linearization Errors} \label{app:lagrange}
 
The linearization errors $\ell_{\mathrm{x},k}$ and $\ell_{\mathrm{y},k}$ in \cref{eq:taylor} depend on the unknown deviation $[x_{\Delta,k}^\top~u_{\Delta,k}^\top]^\top=:p_k$ from the reference trajectory, but are bounded once $p_k$ is confined to a set $\mathcal{P}_k$. By applying the conservative linearization of \cite[Sec.~IV]{althoff2008nonlin},~\cite[Sec.~3.1]{althoff2013nonlin} to our discrete-time setting, we bound the deviations by the reachable state set and the input uncertainty set as
\begin{align}\label{eq:deviationSet}
    \mathcal{P}_k = \bigl(\widehat{\mathcal{X}}_k - \bar{x}_k\bigr) \times \mathcal{U}_{\Delta,k},
\end{align}
where $\widehat{\X}_{0} = \{\bar{x}_{0}\}\oplus\XO$ and the overapproximative reachable state set at $k>0$ follows by projecting $\widehat{\Z}_k$ of \cref{thm:jointReach} onto the state block, i.e., $\widehat{\X}_{k} = \Px\widehat{\Z}_{k-1}$. As the linearization errors vanish for $p_k=\mathbf{0}$ and $\mathbf{0}\in\mathcal{P}_0$, induction over $k$ yields $\bar{x}_k\in\widehat{\X}_k$, i.e., $\mathbf{0}\in\mathcal{P}_k$.
 
Given the set $\mathcal{P}_k$, the bounds for the linearization errors $\ell_{\mathrm{x},k}$ and $\ell_{\mathrm{y},k}$ can be computed by evaluating the Taylor expansion with Lagrange remainder over $\mathcal{P}_k$.
For a scalar-valued function $h$ of the combined state-input vector $\tilde{p}$, the linearization errors attainable over $\mathcal{P}_k$ are contained in the set
\begin{align}\label{eq:remainderSet}
    \mathcal{L}&_k(h) := \biggl\{
        \sum_{j=2}^{\kappa}
        \frac{(p_k^\top \nabla)^j h(\tilde{p})}{j!}
        \Big|_{\tilde{p} = \bar{p}_k} \notag \\
        &+ \frac{(p_k^\top \nabla)^{\kappa+1} h(\tilde{p})}{(\kappa+1)!}
        \Big|_{\tilde{p} \in \bar{p}_k \oplus \mathcal{P}_k}
        \bigg|\;
        p_k \in \mathcal{P}_k
    \biggr\},
\end{align}
where $(p_k^\top \nabla)^j h$ is the $j$-order directional derivative of $h$ in the direction $p_k$ and $\bar{p}_k = [\bar{x}_k^{\top}~\bar{u}_k^{\top}]^\top$ is the reference state-input pair of \cref{eq:refTraj}. The sum collects the Taylor terms of order $2,\dots,\kappa$ and the second term is the Lagrange remainder of order $\kappa+1$~\cite{Berz1998}, where $\kappa\geq 1$ is user-defined; the exact remainder is evaluated on the segment from $\bar{p}_k$ to $\bar{p}_k+p_k$, which lies in the convex set $\bar{p}_k\oplus\mathcal{P}_k$ that \cref{eq:remainderSet} uses to decouple $\tilde{p}$ from $p_k$. For $\kappa=1$ the sum is empty and \cref{eq:remainderSet} reduces to the second-order Lagrange remainder, so that the framework requires no more than $f,g\in C^2$; larger $\kappa$ reduces the overestimation error but requires $f,g\in C^{\kappa+1}$. Evaluating \cref{eq:remainderSet} for each component of $f$ and $g$ by interval arithmetic~\cite{jaulin2001interval} leads to intervals $\widehat{\mathcal{L}}_{\mathrm{x},k(i)}\supseteq\mathcal{L}_k(f_{(i)})$ and $\widehat{\mathcal{L}}_{\mathrm{y},k(i)}\supseteq\mathcal{L}_k(g_{(i)})$, whose Cartesian products are the sought enclosures
\begin{subequations}\label{eq:LHat}
\begin{align}
    \widehat{\mathcal{L}}_{\mathrm{x},k} &= \widehat{\mathcal{L}}_{\mathrm{x},k(1)}\times\dots\times\widehat{\mathcal{L}}_{\mathrm{x},k(\nx)}, \label{eq:LHatX}\\
    \widehat{\mathcal{L}}_{\mathrm{y},k} &= \widehat{\mathcal{L}}_{\mathrm{y},k(1)}\times\dots\times\widehat{\mathcal{L}}_{\mathrm{y},k(\ny)}. \label{eq:LHatY}
\end{align}
\end{subequations}
The connection between $\mathcal{P}_k$ and the linearization errors is not circular, since $\widehat{\X}_k$ only depends on the sets $\widehat{\mathcal{L}}_{\mathrm{x},i}$ and $\mathcal{P}_{i}$ of the steps $i=0,\dots,k-1$, so that the enclosures can be constructed step by step.
 
For $\mathsf{R}=\mathrm{sim}$, \cref{lem:simTraj} requires the uncertainty sets of every test case $m$ to contain those of its one-step test cases $(m,k)$.
\begin{lemma}[Enclosures for simulation conformance]\label{lem:simReq}
Let $\sysM$ be a system as in \cref{def:zds} with $\ax=\mathbf{0}$, $\bar{x}_0^{(m)}=x_0^{(m)}(\sysT)$ for all $m\in\M$, and the uncertainty sets $\mathcal{E}_{\mathrm{x},k}=-\widehat{\mathcal{L}}_{\mathrm{x},k}$ and $\mathcal{E}_{\mathrm{y},k}=-\widehat{\mathcal{L}}_{\mathrm{y},k}$, using the linearization error enclosures in \cref{eq:LHat} with $\kappa=1$. If $\sysT\preceq_{\mathrm{sim}(\M)}\sysM$, then \cref{eq:simReq} holds.
\end{lemma}
\vspace{0.1cm}\textit{Proof:} See Appendix~\ref{app:proofSimReq}.\vspace{0.2cm}

\section{Proofs} \label{app:proofs}
 
\subsection{Proof of Lemma~\ref{lem:simTraj}}\label{app:proofSimTraj}
Since $\ax=\mathbf{0}$, the one-step test case $(m,k)$ starts in $x_k^{(m)}(\sysT)$, so that by \cref{def:simConfData} there exist $u_k\in\bar{u}_k^{(m)}\oplus\mathcal{U}_{\Delta,k}^{(m)}$, $e_{\mathrm{x},k}\in\mathcal{E}_{\mathrm{x},0}^{(m,k)}$, and $e_{\mathrm{y},k}\in\mathcal{E}_{\mathrm{y},0}^{(m,k)}$ for which \cref{eq:sysNL} maps $x_k^{(m)}(\sysT)$ to $x_{k+1}^{(m)}(\sysT)$ and $y_k^{(m)}(\sysT)$. By \cref{eq:simReq}, these choices are also admissible at step $k$ of test case $m$, which has the same nominal input and input uncertainty set. Since \cref{eq:sysNL} depends only on the current state and the choices at the current step, these one-step transitions can be concatenated: applying them for $k=0,\dots,\nk-1$ from $x_0^{(m)}=\bar{x}_0^{(m)}=x_0^{(m)}(\sysT)$ yields a trace of $\sysM$ for test case $m$ that attains all measured states and outputs, i.e., $z_{0:\nk-1}^{(m)}(\sysT)\in\mathcal{T}_{\mathrm{z}}^{(m)}$.

\subsection{Proof of Theorem~\ref{thm:jointReach}}\label{app:proofJointReach}
Both inclusions follow from \cref{eq:linSolution}, which describes the traces of $\sysM$ exactly for some $\ell_{\mathrm{x},i}\in{\mathcal{L}}_{\mathrm{x},i}$ and $\ell_{\mathrm{y},k}\in{\mathcal{L}}_{\mathrm{y},k}$, due to the conservative linearization of \cite[Sec.~IV]{althoff2008nonlin}.
 
\emph{Overapproximation:} Replacing every uncertain variable in \cref{eq:linSolution} by its bounding set, i.e., $x_{\Delta,0}$ by $\XO$, $u_{\Delta,i}$ by $\mathcal{U}_{\Delta,i}$, $e_{\mathrm{x},i}$ by $\mathcal{E}_{\mathrm{x},i}$, $e_{\mathrm{y},k}$ by $\mathcal{E}_{\mathrm{y},k}$, $\ell_{\mathrm{x},i}$ by $\widehat{\mathcal{L}}_{\mathrm{x},i}\supseteq\mathcal{L}_{\mathrm{x},i}$, and $\ell_{\mathrm{y},k}$ by $\widehat{\mathcal{L}}_{\mathrm{y},k}\supseteq\mathcal{L}_{\mathrm{y},k}$, and the sums by Minkowski sums leads to the overapproximation $\widehat{\Z}_k$ in \cref{eq:ZHat}.
 
\emph{Underapproximation:} Our construction reverses the Minkowski-difference inner approximation of~\cite[Prop.~1]{wetzlinger2024inner}: the model uncertainties enlarge the reachable set of $\sysM$ such that subtracting the linearization errors yields $\bar{\Z}_k$. Let $\bar{z}\in\bar{\Z}_k$, i.e., $\bar{z}=\bar{z}_k+\Apropm{k}x_{\Delta,0}+\sum_{i=0}^{k}\Bpropm{k}{i}u_{\Delta,i}$ for some $x_{\Delta,0}\in\XO$ and $u_{\Delta,i}\in\mathcal{U}_{\Delta,i}$ by \cref{eq:ZBar}. Choosing step by step $e_{\mathrm{x},i}=-\ell_{\mathrm{x},i}$ and $e_{\mathrm{y},k}=-\ell_{\mathrm{y},k}$, which \cref{eq:uncChoice} admits since $\ell_{\mathrm{x},i}\in\widehat{\mathcal{L}}_{\mathrm{x},i}$ and $\ell_{\mathrm{y},k}\in\widehat{\mathcal{L}}_{\mathrm{y},k}$, and inserting them into \cref{eq:linSolution} cancels the terms $\Cpropm{k}{i}(e_{\mathrm{x},i}+\ell_{\mathrm{x},i})$ and $\Dprop(e_{\mathrm{y},k}+\ell_{\mathrm{y},k})$, so that $z_k=\bar{z}$, which proves \cref{eq:Zunder}.
 
\subsection{Proof of Theorem~\ref{thm:jointTrace}}\label{app:proofJointTrace}
Omitting the terms $e_{\mathrm{x},i}+\ell_{\mathrm{x},i}$ and $e_{\mathrm{y},k}+\ell_{\mathrm{y},k}$ in \cref{eq:linSolution} and stacking the result for $k=0,\dots,\nk-1$ leads to the traces ${z}_{0:\nk-1}=\bar{z}_{0:\nk-1}+\ATpropm x_{\Delta,0}+\BTpropm u_{\Delta,0:\nk-1}$, with $\ATpropm$ and $\BTpropm$ in \cref{eq:propMatTrace}; ranging over all $x_{\Delta,0}\in\XO$ and $u_{\Delta,0:\nk-1}\in\mathcal{U}_{\Delta,0:\nk-1}$ yields \cref{eq:TzBar}. Each such trace is a trace of $\sysM$: the choice $e_{\mathrm{x},i}=-\ell_{\mathrm{x},i}$, $e_{\mathrm{y},i}=-\ell_{\mathrm{y},i}$, which \cref{eq:uncChoice} admits, depends only on the current state $x_i$ and input $u_i$, so it is a single realization for all $k$, and inserting it into \cref{eq:linSolution} cancels the omitted terms. This proves \cref{eq:Tunder}, where other admissible choices generally yield further traces of $\sysM$.
 
\subsection{Proof of Corollary~\ref{cor:containment}}\label{app:proofContainment}
For $\mathsf{R}=\mathrm{reach}$, inserting \cref{eq:YBar} into \cref{eq:reachContainment} and applying \cref{eq:Zunder} leads to $y_k(\sysT)\in\Py\bar{\Z}_k\subseteq\Py\Z_k=\Y_k$, since linear maps preserve inclusions, i.e., $\sysT\preceq_{\mathrm{reach}(\M)}\sysM$ by \cref{def:reachConfData}. For $\mathsf{R}=\mathrm{trace}$, inserting \cref{eq:TyBar} into \cref{eq:traceContainment} and applying \cref{eq:Tunder} leads to $y_{0:\nk-1}(\sysT)\in\PT\bar{\mathcal{T}}_{\mathrm{z}}\subseteq\PT\mathcal{T}_{\mathrm{z}}=\mathcal{T}_{\mathrm{y}}$, i.e., $\sysT\preceq_{\mathrm{trace}(\M)}\sysM$ by \cref{def:traceConfData}. For $\mathsf{R}=\mathrm{sim}$, \cref{eq:ZBar} for the one-step test case $(m,k)$ reduces to \cref{eq:ZBarSim}, since $\ax=\mathbf{0}$ and $\mathcal{U}_{\Delta,0}^{(m,k)}=\mathcal{U}_{\Delta,k}^{(m)}$. Inserting \cref{eq:ZBarSim} into \cref{eq:simContainment} and applying \cref{eq:Zunder} to $(m,k)$ leads to $z_k^{(m)}(\sysT)\in\bar{\Z}_0^{(m,k)}\subseteq\Z_0^{(m,k)}$, i.e., $\sysT\preceq_{\mathrm{sim}(\M)}\sysM$ by \cref{def:simConfData}.
 
\subsection{Proof of Proposition~\ref{prop:constrAll}}\label{app:proofConstrAll}
\sloppy We give the chain for $\mathsf{R}=\mathrm{trace}$ and then the changes for the other two relations:
\begin{align*}
&y_{0:\nk-1}(\sysT)\in\bar{\mathcal{T}}_{\mathrm{y}}
\\
\overset{\text{\cref{eq:TzBarAll},\cref{eq:TyBar}}}{\Leftrightarrow}\,&\exists x_{\Delta,0}\in\XO,\,u_{\Delta,i}\in\mathcal{U}_{\Delta,i},\,i=0,\dots,\nk-1\colon \\
&\quad y_{k}(\sysT) = \bar{y}_k + \Py\Apropm{k} x_{\Delta,0}\\&\qquad + \sum_{i=0}^k \Py\Bpropm{k}{i} u_{\Delta,i},~ k=0,\dots,\nk-1
\\
\overset{\text{\cref{eq:uncSetsStructure},\cref{eq:cont_gen}}}{\Leftrightarrow}\,&\exists \lambx \in \mathbb{R}^{\nux},\, \lambu{0},\dots, \lambu{\nk-1} \in \mathbb{R}^{\nuu}\colon \\
    &\quad |\lambx| \leq \textbf{1},~|\lambu{0}| \leq \textbf{1},\dots,~|\lambu{\nk-1}| \leq \textbf{1},  \\
    &\quad x_{\Delta,0} = \Gx \mathrm{diag}(\ax)\lambx,\\
    &\quad
    u_{\Delta,i} = \Gu \mathrm{diag}(\au)\lambu{i},~ i=0,\dots,\nk-1,\\
    &\quad y_{k}(\sysT) = \bar{y}_k + \Py\Apropm{k}x_{\Delta,0}\\&\qquad + \sum_{i=0}^k \Py\Bpropm{k}{i}u_{\Delta,i},~ k=0,\dots,\nk-1
\\
\Leftrightarrow\,&\exists \bx \in \mathbb{R}^{\nux},\, \bu{0},\dots,\bu{\nk-1} \in \mathbb{R}^{\nuu}\colon \\
    &\quad |\bx| \leq \ax,\, |\bu{0}| \leq \au, \dots,\,
    |\bu{\nk-1}| \leq \au, \\
    &\quad y_{\Delta,k} = \Py\Apropm{k} \Gx \bx\\&\qquad + \sum_{i=0}^k \Py\Bpropm{k}{i} \Gu \bu{i},~ k=0,\dots,\nk-1,
\end{align*}
where the last step substitutes $\bx:=\mathrm{diag}(\ax)\lambx$ and $\bu{i}:=\mathrm{diag}(\au)\lambu{i}$, which turns $|\lambda|\leq\mathbf{1}$ into $|\beta|\leq\alpha$ since $\alpha\geq\mathbf{0}$, and moves $\bar{y}_k$ to the left-hand side. Stacking the bounds and the equalities for $k=0,\dots,\nk-1$ with $\beta=[\bx^{\top}~\bu{0}^{\top}~\dots~\bu{\nk-1}^{\top}]^\top$ leads to \cref{eq:constrAll} with the variables in \cref{tab:constrMatrices}. For $\mathsf{R}=\mathrm{reach}$, applying the same chain to each time step with its own realization (cf.~\cite[Lemma~4]{luetzow2026reachset}) leads to one block $\beta_k$ per step and thus the block-diagonal $M_{\mathrm{reach}}$. For $\mathsf{R}=\mathrm{sim}$, applying it to each one-step test case $(m,k)$ with $x_{\Delta,0}=\mathbf{0}$, the full deviation $z_{\Delta,k}$ in place of $y_{\Delta,k}$, and $\Bprop{0}{0}{m,k}$ in place of $\Py\Bpropm{k}{i}$ leads to the variables in \cref{tab:constrMatrices}, in which the single variable $\bu{k}$ enters both the state and the output block of $z_{\Delta,k}$, retaining the coupling of the successor state and the output.
 
\subsection{Proof of Lemma~\ref{lem:cost}}\label{app:proofCost}
Inserting \cref{eq:uncSetsStructure} into \cref{eq:ZBar,eq:YBar} leads to the generators $\Py\Apropm{k}\Gx\mathrm{diag}(\ax)$ and $\Py\Bpropm{k}{j}\Gu\mathrm{diag}(\au)$, $j=0,\dots,k$, of $\bar{\Y}_k$. Inserting them into \cref{def:intNorm} and \cref{eq:cost} and using $|V_iG\,\mathrm{diag}(a)|\mathbf{1}=|V_iG|\,a$ for $a\geq\mathbf{0}$ leads to \cref{eq:objective} with \cref{eq:costCoeff} and $\alpha=[\ax^\top~\au^\top]^\top$ (cf.~\cite[Lemma~2]{luetzow2026reachset}, \cite[Eq.~(14)]{luetzow2025zonoconformal}).

\subsection{Proof of Theorem~\ref{thm:confId}}\label{app:proofConfId}
The optimizer $(\alpha^\ast(\M),\beta^\ast(\M))$ satisfies \cref{eq:confConstr}, i.e., \cref{eq:constrAll} for all $m\in\M$, which by \cref{prop:constrAll} leads to \cref{eq:containment} for all $m\in\M$ and thus $\sysT\preceq_{\mathsf{R}(\M)}\sysM$ by \cref{cor:containment}.
 
\subsection{Proof of Lemma~\ref{lem:projection}}\label{app:proofProjection}
For each test case $m$, \cref{eq:constrAll} consists of linear equalities and inequalities in $(\alpha,\beta^{(m)})$ whose data-dependent coefficients $d_{\mathsf{R}}^{(m)}$ and $M_{\mathsf{R}}^{(m)}$ are those of test case $m$. Projecting this polyhedron onto $\alpha$ leads to $\mathcal{A}_{\mathsf{R}}^{(m)}$, which is again a closed convex polyhedron by Fourier--Motzkin elimination~\cite[Sec.~12.2]{schrijver1986theory}. Since each $\beta^{(m)}$ occurs only in the constraints of test case $m$, projecting the feasible set of \cref{eq:confOpt} onto $\alpha$ leads to the feasible set of \cref{eq:projProgram}; since $\gamma\alpha$ does not depend on $\beta$, both programs have the same optimal value and the same optimal $\alpha$.
 
\subsection{Proof of Theorem~\ref{thm:guarantees}}\label{app:proofGuarantees}
By \cref{lem:projection}, \cref{eq:confOpt} is equivalent to the convex program \cref{eq:projProgram} with decision variable $\alpha\in\mathbb{R}^{\ntheta}$, where $\ntheta=\nuu$ for $\mathsf{R}=\mathrm{sim}$ since $\ax=\mathbf{0}$, and one constraint set $\mathcal{A}_{\mathsf{R}}^{(m)}$ per test case, which depends only on the data of $m$. By \cref{ass:independence}, the sets $\mathcal{A}_{\mathsf{R}}^{(m)}$, $m\in\M$, and $\mathcal{A}_{\mathsf{R}}^{(\tilde{m})}$ are i.i.d., so \cref{eq:projProgram} is a convex scenario program with at most $\ntheta$ support constraints~\cite[Thm.~2]{calafiore2005uncertain}. Under \cref{ass:nondegen}, applying \cite[Eq.~(14) and (15)]{osullivan2025bridgingconformalpredictionscenario} with zero outliers bounds $\mathbb{P}\{\alpha^\ast(\M)\in\mathcal{A}_{\mathsf{R}}^{(\tilde{m})}\mid\M\}$ from below as in \cref{eq:expectedCov,eq:varCov}. By \cref{prop:constrAll,cor:containment}, $\alpha^\ast(\M)\in\mathcal{A}_{\mathsf{R}}^{(\tilde{m})}$ implies \cref{eq:containment} for $\tilde{m}$ and thus $\sysT\preceq_{\mathsf{R}(\tilde{m})}\sysM$, so that $\eta(\M)\geq\mathbb{P}\{\alpha^\ast(\M)\in\mathcal{A}_{\mathsf{R}}^{(\tilde{m})}\mid\M\}$ and both bounds carry over.
 
\subsection{Proof of Lemma~\ref{lem:simReq}}\label{app:proofSimReq}
We first prove $x_k^{(m)}(\sysT)\in\widehat{\X}_k^{(m)}$ by induction, assuming that it implies \cref{eq:simReq} at step $k$. For $k=0$, $\widehat{\X}_0^{(m)}=\{\bar{x}_0^{(m)}\}=\{x_0^{(m)}(\sysT)\}$ since $\ax=\mathbf{0}$. If the inclusion, and thus \cref{eq:simReq}, holds up to step $k$, concatenating the one-step transitions as in the proof of \cref{lem:simTraj} attains $x_{k+1}^{(m)}(\sysT)$ under $\sysM$, so that $x_{k+1}^{(m)}(\sysT)\in\widehat{\X}_{k+1}^{(m)}$ by \cref{thm:jointReach}. It remains to show that $x_k^{(m)}(\sysT)\in\widehat{\X}_k^{(m)}$ implies \cref{eq:simReq} at step $k$. For $\kappa=1$, the enclosures $\widehat{\mathcal{L}}_{0}^{(m,k)}$ of the one-step test case $(m,k)$ and $\widehat{\mathcal{L}}_{k}^{(m)}$ of test case $m$ evaluate the same expression \cref{eq:remainderSet}, but over different sets: as $\ax=\mathbf{0}$, the one-step test case has $\mathcal{P}_0^{(m,k)}=\{\mathbf{0}\}\times\mathcal{U}_{\Delta,k}^{(m)}$ and $\bar{p}_0^{(m,k)}=[x_k^{(m)}(\sysT)^\top~\bar{u}_k^{(m)\top}]^\top$. Since $\mathbf{0}\in\widehat{\X}_k^{(m)}-\bar{x}_k^{(m)}$ and $x_k^{(m)}(\sysT)\in\widehat{\X}_k^{(m)}$, it follows that $\mathcal{P}_0^{(m,k)}\subseteq\mathcal{P}_k^{(m)}$ and $\bar{p}_0^{(m,k)}\oplus\mathcal{P}_0^{(m,k)}\subseteq\bar{p}_k^{(m)}\oplus\mathcal{P}_k^{(m)}$, so that inclusion monotonicity~\cite[Sec.~4.3]{moore2009introduction} yields $\widehat{\mathcal{L}}_{0}^{(m,k)}\subseteq\widehat{\mathcal{L}}_{k}^{(m)}$ and thus \cref{eq:simReq}.
 
\section*{Use of Generative AI}
Claude Code~\cite{anthropic2026claudecode} was used to improve the language of the article and to implement the methods and experiments; all technical content, results, and conclusions are the authors’ own, and the authors take full responsibility for them.

\section*{References}\vspace{-0.4cm}
\bibliographystyle{ieeetr}
\bibliography{literature}

\begin{IEEEbiography}[{\includegraphics[width=1in,height=1.25in,clip,keepaspectratio]{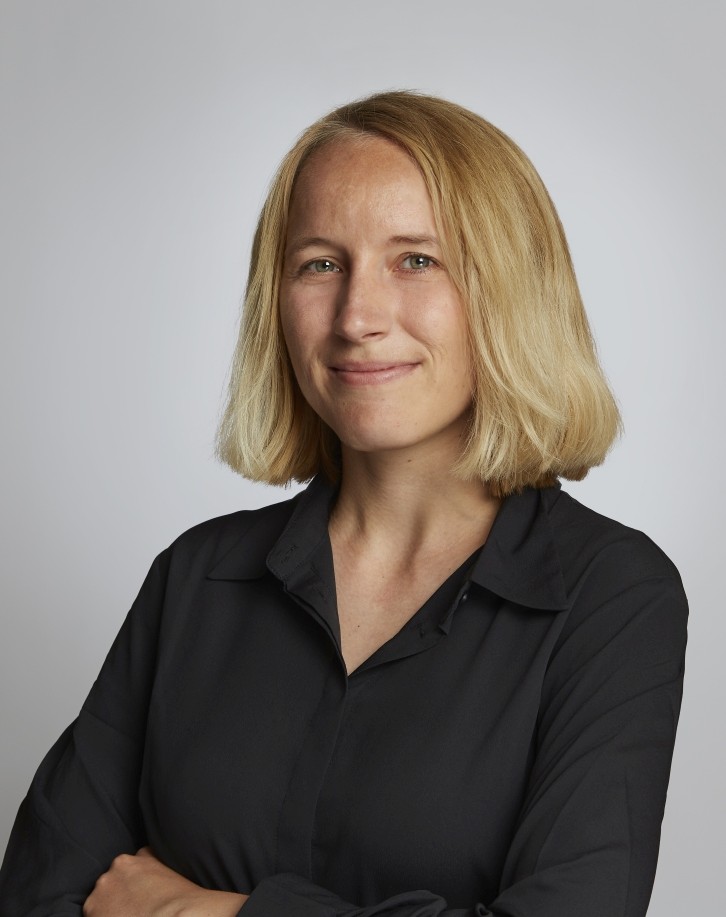}}]{Laura Lützow} received the Bachelor of Science degree in mechatronics from the Technical University of Ilmenau, Germany, in 2020, and the Master of Science degree in robotics, cognition and intelligence and the Ph.D. degree in computer science both from the Technical University of Munich, Germany, in 2022 and 2026, respectively.
In 2022 and 2025, she was a visiting student researcher at
the Massachusetts Institute of Technology, USA, and at Stanford University, USA, respectively.
She is currently working as a Postdoctoral Researcher at the Stanford Intelligent Systems Laboratory, Stanford University, USA.
Her research interests include formal methods, system identification, control theory, and uncertainty quantification, with applications to safety-critical systems.
\end{IEEEbiography}

\begin{IEEEbiography}[{\includegraphics[width=1in,height=1.25in,clip,keepaspectratio]{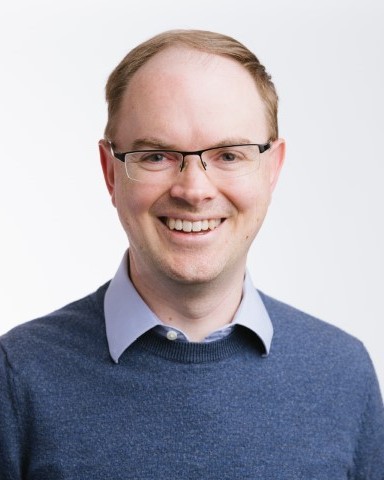}}]
{Mykel J. Kochenderfer} (Senior Member, IEEE) received B.S. and M.S. degrees in computer science from Stanford University in 2003 and the Ph.D. degree in informatics from the University of Edinburgh in 2006. From 2006 to 2013, he was a member of the technical staff at MIT Lincoln Laboratory. In 2013, he joined Stanford University as a professor in the Department of Aeronautics and Astronautics and (by courtesy) Computer Science. He directs the Stanford Intelligent Systems Laboratory and co-directs the Stanford Center for AI Safety. His research focuses on advanced algorithms and analytical methods for the design of robust decision making systems. He is the author of several textbooks spanning decision making, optimization, and validation.
\end{IEEEbiography}
 
\begin{IEEEbiography}
[{\includegraphics[width=1in,height=1.25in,clip,keepaspectratio]{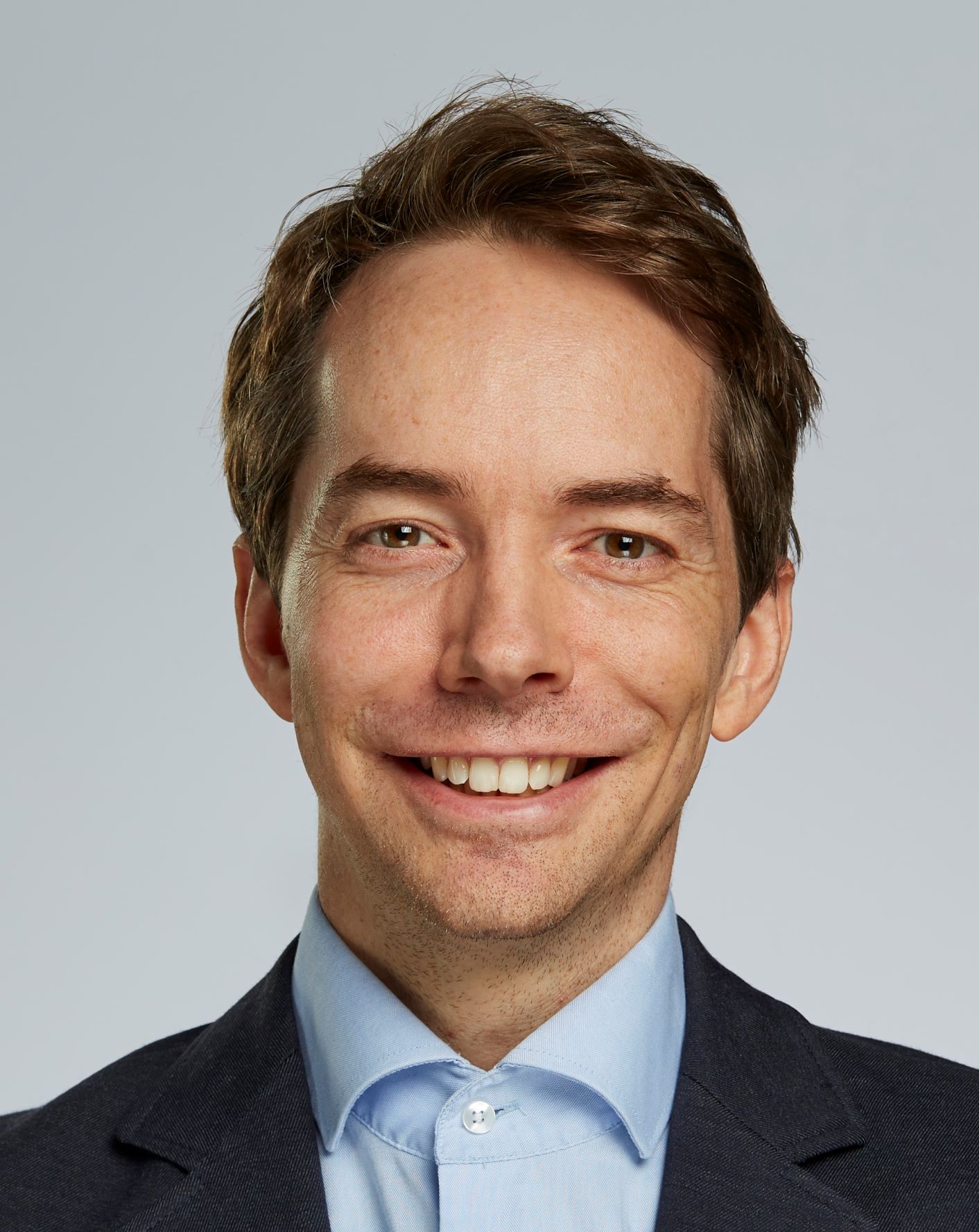}}]{Matthias Althoff} (Member, IEEE)
received the Diploma Engineering degree in mechanical engineering and the Ph.D. degree in electrical engineering, both from the Technical University of Munich, Germany, in 2005 and 2010, respectively.
From 2010 to 2012, he was a Postdoctoral Researcher at Carnegie Mellon University, Pittsburgh, PA, USA, and from 2012 to 2013, he was an Assistant Professor at the Technical University of Ilmenau, Germany.
He is currently an Associate Professor in computer science with the Technical University of Munich. His research interests include formal verification of continuous and hybrid systems, reachability analysis, planning algorithms, nonlinear control, robotics, automated vehicles, and power systems.
\end{IEEEbiography}

\end{document}